\magnification1200


\vskip 2cm
\centerline
{\bf   E theory in seven dimensions}
\vskip 1cm
\centerline{ Michaella Pettit and Peter West}
\centerline{Department of Mathematics}
\centerline{King's College, London WC2R 2LS, UK}
\vskip 2cm
\leftline{\sl Abstract}
We construct the non-linear realisation of the semi-direct product of $E_{11}$ and its vector representation in its decomposition into the subalgebra $GL(7)\otimes SL(5)$ to find a seven dimensional theory. The resulting equations of motion essentially follow from the Dynkin diagram of $E_{11}$ and if one restricts them to contain only the usual fields of supergravity and the derivatives with respect to the usual coordinates of spacetime then these are the equations of motion of seven dimensional supergravity.

\vskip2cm
\noindent

\vskip .5cm

\vfill
\eject
{\bf 1 Introduction}
\medskip
It has been conjectured [1, 2]  that the low energy effective action of strings and branes is the non-linear realisation of  $E_{11} \otimes_s l_1$, where $\otimes_s$ is the semi-direct product and $l_1$ is the vector representation.  The fields of the theory arise out of the $E_{11}$ algebra, and the generalised space-time that the fields depend on is  appears due to  the $l_1$ representation. The different maximal supergravity theories appear when one takes different decompositions of $E_{11}$ [3,4,5,6].  For each decomposition the fields and coordinates are classified according to a level and the low level fields are those of the corresponding maximal supergravity and the level zero coordinates are the usual coordinates of spacetime. The equations of motion follow from the symmetries of the non-linear realisation
and it was clear from early on that the $E_{11} \otimes_s l_1$ non-linear realisation might  contain all the maximal supergravities in a single theory. Although the early $E_{11}$ papers contain part of the equations of motion in various dimensions it was only in references [7,8] that
the equations of motion were found for the eleven and five dimensional theories. When these equations, which essentially followed uniquely from the non-linear realisation, were truncated to contain only the fields that are associated with the supergravity theories and the only coordinates is the usual coordinates of spacetime, then the equations are precisely with those of the eleven and five dimensional supergravity theories. For a review of E theory see reference [9].
\par
In this paper we will carry out the decomposition of the $E_{11} \otimes_s l_1$  algebra into $GL(7)\otimes GL(5)$ to find the seven dimensional theory and then construct the corresponding non-linear realisation and so the equations of motion in seven dimensions.
In more detail in section 2, we derive the decomposition of $E_{11} \otimes l_1$ corresponding to the seven dimensional theory   and then use these results  to construct the Cartan forms in section 3.  Section 4 focuses on finding the transformations of the Cartan forms and in section 5 we  derive the equations of motion of seven  dimensional theory.
\par
We first recall the non-linear realisation, specifically with respect to the algebra we are interested in, namely,  $E_{11} \otimes_s  l_1$. This has been discussed  in previous papers, but we will briefly  repeat it here for convenience. We construct the non-linear realisation using the group element $g \in E_{11} \otimes_s l_1$ with
$$ g = g_lg_E \; ,
\eqno(1.1) $$
where $g_l$ is the group element made of generators of the $l_1$ representation of the $E_{11}$ algebra, and  $g_E$ is a group element of $E_{11}$. In terms of the generators, these group elements can be written as
$$ g_l = e^{z^A l_A}  \; , \;\;\;  g_E = e^{A_{\underline \alpha} R^{\underline \alpha}} \; .
\eqno(1.2) $$
where $ R^{\underline \alpha}$ are the generators of the $E_{11}$ algebra and $l_A$ are the generators of the vector representation of $E_{11}$. The $A_{\underline \alpha}$ will turn out to  be the fields of our theory, and the $z^A$ are the generalised space-time coordinates upon which  the fields will depend.
\par
The non-linear realisation is invariant under transformations
$$ g \rightarrow g_0g \, , \;\;\; g_0 \in E_{11} \otimes_s l_1 \, , \eqno(1.3) $$
$$ g \rightarrow gh \, , \;\;\; g \in I_c(E_{11}) \, . \eqno(1.4) $$
where $g_0 \in E_{11} $  is  a rigid transformation, and $h $ is an element of the Cartan involution invariant subalgebra $I_c(E_{11}) $, and is a local transformation. The Cartan involution acts on a generator of $E_{11}$ as $ I_c(R^{\underline \alpha}) = - R^{- \underline \alpha} $ for a root $ {\underline \alpha}$, and hence the Cartan involution invariant subalgebra is generated by $ R^{\underline \alpha} - R^{-\underline \alpha} $.
\par
The dynamics of the non-linear realisation are a set of equations of motion  which is invariant under the transformations in equation (1.3)-(1.4). We will construct the invariant dynamical equations from  the Cartan forms which are given by
$$ {\cal V} = g^{-1}dg = {\cal V}_E + {\cal V}_l \, ,
\eqno(1.5) $$
which we have split into terms containing the $E_{11}$ generators and terms containing the $l_1$ representation, explicitly
$$ {\cal V}_E = g^{-1}_Edg_E = dz^{\Pi}G_{\Pi, \underline \alpha}R^{\underline \alpha} \, ,
\eqno(1.6) $$
$$ {\cal V}_l = g^{-1}_ldg_l = g_E^{-1}(g_l^{-1}dg_l)g_E = dz^{\Pi}E_{\Pi}{}^Al_A \, ,
 \eqno(1.7) $$
where $G_{\Pi, \underline \alpha}$ are the Cartan forms of $E_{11}$ and $E_{\Pi}{}^A = (e^{A_{\underline \alpha}D^{\underline \alpha}})_{\Pi}{}^A$ is the vielbein on the generalised spacetime.
\par
Both ${\cal V}_E$ and ${\cal V}_l$ are invariant under the rigid transformations, but  transform in the following way under the local transformations
$$ {\cal V}_E \rightarrow h^{-1}{\cal V}_E h + h^{-1}dh \, , \;\;\;  {\cal V}_l \rightarrow h^{-1}{\cal V}_l h \, . \eqno(1.8) $$
\medskip
{\bf 2  The $E_{11}$ algebra in seven dimensions}
\medskip
We begin by giving the algebra of $E_{11}$ in 7 dimensions. The algebra of the positive root generators was given in the paper  of reference [10].  In this paper we will extend these to the full $E_{11}$ algebra and so include the negative root generators. We will also construct the commutators of the $E_{11}$ generators with those of the $l_1$ representation.  Furthermore, we will construct the Cartan involution invariant subalgebra, dentoed by $ I_c(E_{11})$ algebra.
\medskip
{\bf 2.1  The $E_{11}$ algebra in seven dimensions}
\medskip
To find the seven dimensional theory we delete node seven in the $E_{11}$ Dynkin diagram, whereupon   we find the algebra
$SL(7)\otimes SL(5)$. The Dynkin diagram is

$$ \matrix{ & & & & & & & & & & & & & & & \bullet & 11 & & & \cr
& & & & & & & & & & & & & & & | & & & & \cr
& \bullet & - & \bullet & - & \bullet & - & \bullet & - & \bullet & - & \bullet & - & \otimes & - & \bullet & - & \bullet & - & \bullet \cr
& 1 & & 2 & & 3 & & 4 & & 5 & & 6 & & 7 & & 8 & & 9 & & 10 \cr} $$

\noindent
The cross on node 7 represents the fact that this is the deleted node.
 Decomposing the $E_{11}$ algbra into representations of this algebra the  generators of level zero and above are given by [10]
$$ K^a{}_b, \ \ R^M{}_N; \ \ R^{aMN} ; \ \ R^{a_1a_2}{}_M; \ \ R^{a_1a_2a_3M}; \ \ R^{a_1\ldots a_4}{}_{MN} ; \ \ R^{a_1\ldots a_5M}{}_N,
$$
$$
  R^{a_1\ldots a_4,b};  \ \ R^{a_1\ldots a_6}{}_{MN,P} , \ \ R^{a_1\ldots a_6}{}^{(MN)},  \ \ R^{a_1\ldots a_5,b}{}^{ MN} , \ldots \eqno(2.1.1)$$
The indices with the round brakets surrounding them are symmetric in their permutation. The indices in the remaining blocks are totally antisymmetric (where a comma indicates a new antisymmetric block), and they belong to irreducible representations of $SL(7)\otimes SL(5)$ and so
$$
\sum_N R^N{}_N=0; \ \  \sum _N R^{a_1\ldots a_5 N}{}_N=0 , \ \ R^{[a_1 \ldots a_4, b]}=0 ;
\eqno(2.1.2)$$
$$
 \ \ R^{[a_1\ldots a_5,b]}{}^{ MN} =0 ;\ \
R^{a_1\ldots a_6}{}_{[M N,P]}=0,\ \
\eqno(2.1.3)$$

\noindent
The generators can be given a level, which in 7 dimensions, is the number of up indices minus the number of down SL(7)  indices, and the semi-colons between the generators in equation (2.1.1) represent an increase in the level.
\par
We begin by explaining how the algebra in seven dimensions was derived.
In general, the algebra of $E_{11}\otimes_s l_1$ can be written
$$ [R^{\alpha}, R^{\beta}] = f^{\alpha\beta}{}_{\gamma}R^{\gamma} \, ,
\eqno(2.1.4)$$
$$ [R^{\alpha},l_A] = -(D^{\alpha})_A{}^Bl_B \, ,
\eqno(2.1.5) $$
where $ (D^{\alpha})_A{}^B $ is the matrix of the first fundamental (vector) representation of $E_{11}$ which satisfies
$$ [D^{\alpha}, D^{\beta}] = f^{\alpha \beta}{}_{\gamma}D^{\gamma} \, .
\eqno(2.1.6) $$

We note that the commutators preserve the level and additionally are preserved under the action of the Cartan involution
$$
I_c(R^{\alpha })= (-1)^{{\rm level}}  R^{-\alpha } \; ,
\eqno(2.1.7)$$
except for level zero which carries a minus sign. This will provide a useful check in  the following derivation of the algebra.

 The generators of the GL(7) algebra are denoted by $K^a{}_b$, $a,b = 1,\ldots , 7 $ satisfy the commutator
 $$ [K^a{}_b, K^c{}_d] = \delta^c_b K^a{}_d - \delta^a{}_d K^c{}_b \, ,
\eqno(2.1.8) $$
 and similarly, the generators of the SL(5) algebra, $R^M{}_N$,  satisfy
 $$ [R^M{}_N, R^P{}_Q] = \delta^P_N R^M{}_Q - \delta ^M_Q R^P{}_N \, ,
\eqno(2.1.9) $$
 
 \noindent
We choose the remaining generators to be irreducible represntations of GL(7) $\times$ SL(5) and hence they satisfy the constraints in equation (2.1.2). Since the generators are representations of GL(7) and SL(5), the commutators with the $K^a{}_b$ and $R^M{}_N$ generators are determined. For example, we have at level 1,
$$ [K^a{}_b, R^{cMN}] = \delta^c_bR^{aMN}  \; ,
\eqno(2.1.10)$$
and at level -1, we have
$$ [K^a{}_b, R_{cMN}] = -\delta_c^aR_{bMN}  \; ,
\eqno(2.1.11)$$
\noindent
The other level generators follow a similar pattern in terms of how the spacetime generator acts on the upper and lower indices.
\par
The action of the spacetime SL(7) is
$$ [K^a{}_b, K^c{}_d] = \delta^c_bK^a{}_d - \delta^a_dK^c{}_b \; ,
\eqno(2.1.12)$$
$$ [K^a{}_b, R^M{}_N] = 0  \; ,
\eqno(2.1.13)$$
$$ [K^a{}_b, R^{cMN}] = \delta^c_bR^{aMN}  \; ,
\eqno(2.1.14)$$
$$ [K^a{}_b, R^{cd}{}_M] = 2\delta_b^{[c}R^{|a|d]}{}_M  \; ,
\eqno(2.1.15)$$
$$ [K^a{}_b, R^{c_1c_2c_3M}] = 3\delta_b^{[c_1}R^{|a|c_2c_3]M} \; ,
\eqno(2.1.16)$$
$$[K^a{}_b,R^{c_1 \ldots c_4}{}_{PQ}] = 4\delta^{[c_1|}_bR^{a|c_2c_3c_4]}{}_{PQ}        \; ,                  
\eqno(2.1.17)  $$
$$[K^a{}_b,R^{c_1\ldots c_5M}{}_N]= 5\delta^{[c_1|}_bR^{a| c_2 \ldots c_5] M}{}_N \; ,       
 \eqno(2.1.18)  $$
$$ [K^a{}_b, R^{c_1\ldots c_4,d}]= 4\delta^{[c_1|}_bR^{a|c_2c_3c_4],d} - 4\delta^d_bR^{a[c_1c_2c_3,c_4]}         \; .         
 \eqno(2.1.19) $$

\noindent
The action of SL(5)
$$ [R^M{}_N, R^P{}_Q] = \delta^P_NR^M{}_Q - \delta^M_QR^P{}_N  \; ,
\eqno(2.1.20)$$
$$ [R^M{}_N, R^{aPQ}] = 2\delta_N^{[P}R^{a|M|Q]} - {2\over5}\delta^M_NR^{aPQ} \; ,
\eqno(2.1.21)$$
$$ [R^M{}_N, R^{ab}{}_P] = -\delta^M_PR^{ab}{}_N + {1\over5}\delta^M_NR^{ab}{}_P \; ,
\eqno(2.1.22)$$

$$ [R^M{}_N, R^{a_1a_2a_3P}] = \delta^P_NR^{a_1a_2a_3M} - {1\over5}\delta^M_NR^{a_1a_2a_3P}  \; ,
\eqno(2.1.23)$$
$$ [R^M{}_N, R^{a_1\ldots a_4}{}_{PQ}] = -2\delta^{M}_{[P|}R^{a_1\ldots a_4}{}_{N|Q]} + {2\over 5}\delta^M_NR^{a_1\ldots a_4}{}_{PQ} \; ,            
 \eqno(2.1.24) $$
$$
[R^M{}_N,R^{a_1\ldots a_5P}{}_Q] = \delta^P_NR^{a_1\ldots a_5M}{}_Q - \delta^M_QR^{a_1\ldots a_5P}{}_N   \; ,
\eqno(2.1.25)  $$
$$[R^M{}_N, R^{a_1\ldots a_4,b}] = 0  \; ,
\eqno(2.1.26) $$

We will  calculate  the $E_{11}$ algebra up to level $\pm 5$. To find the right hand side of the commutators of two generators we write down all possible generators of the required level in such a way that it transforms in the same way under GL(7) $\times$ SL(5) as the left-hand side of the commutator.  We then use the  Jacobi identities to fix the coefficient in front of relevant generators. As an example, we find that the commutator of the level 1 generator $R^{a MN}$ with itself leads to a generator of level two. Looking at equation (2.1.1),  we find that the  only such candidate is the generator $R^{ab}{}_R$   and so
$$ [R^{a MN}, R^{ b PQ}]\propto  R^{ab}{}_R  \; ,
\eqno(2.1.27) $$
Demanding that  the level two generator occur so as to have the same GL(7) x SL(5) as the left-hand side we find that the only possibillity is given by
$$ [R^{aMN}, R^{bPQ}] = \varepsilon^{MNPQR}R^{ab}{}_R  \; .
\eqno(2.1.28)$$
The coefficient in front of the level two generator can be fixed to be any number and this determines the normalisation with which the level two generator enters the algebra; we took the coefficient to be one.
\par
We notice that up to level 4, there is only one generator at each level and so this is the only generator that can occur in commutators that result in level four generators. At level five there are two generators and both of these can enter the commutator. For example, if we consider the commutator of  the level one generator with the level four generator, we find that their result, taking account that we must have the same  GL(7) $\times$ SL(5) representations on both sides of the equation, must be given by
$$
[ R^{a_1  MN} , R^{a_2 ...a_5}{}_{PQ} ]= -2 \delta^{[M}_{[P}
  R^{a_1 ...a_5  N]}{}_{Q]}+ \delta_{PQ}^{MN} R^{a_2 \ldots a_5, a_1}  \; .
\eqno(2.1.29)$$
We can choose the coefficients in front of the level five generators to be as above as this is the first time we have encountered them and this choice fixes their normalisation. To find the commutator of the level two and level three generators we use the Jacobi identity
$$ [ [ R^{a_1 a_2}{}_M , R^{a_3 a_4}{}_{ N}], R^{a_5PQ}] + { \rm cyclic} = 0 \; ,
\eqno(2.1.30) $$
as well as the previously derived result for the commutator of the two level two generators given below. We find that
$$ [ R^{a_1 a_2}{}_M , R^{a_3 a_4 a_5 N}]  =  R^{a_1 ...a_5 N}{}_M
+2 \delta^N_R R^{a_3a_4a_5[a_1 , a_2]}
\eqno(2.1.31) $$
\par
Using the above procedure the commutators of the positive level generators up to level five are given as follows  [10].  The commutators formed by repeated use of the  level one  generator $R^{aMN}$ are given by
$$ [R^{aMN}, R^{bPQ}] = \varepsilon^{MNPQR}R^{ab}{}_R  \; ,
\eqno(2.1.32)$$
$$ [R^{aMN}, R^{b_1b_2}{}_P] = \delta_P^{[M}R^{ab_1b_2N]}  \; ,
\eqno(2.1.33)$$
$$
[ R^{a_1 a_3}{}_M , R^{a_3 a_4}{}_N ] = R^{a_1 ...a_4 }{}_{MN} \; ,
\eqno(2.1.34)$$
$$
  [ R^{a_1  MN} , R^{a_2 a_3 a_4P}] = \varepsilon^{MNPQR} R^{a_1
  ...a_4}{}_{QR} \; ,
 \eqno(2.1.35)$$
$$
[ R^{a_1 a_2}{}_M , R^{a_3 a_4 a_5  N} ] = R^{a_1 ...a_5 N}{}_M
+2 \delta^N_R R^{a_3a_4a_5[a_1 , a_2]} \; ,
\eqno(2.1.36)$$
$$
[ R^{a_1  MN} , R^{a_2 ...a_5}{}_{PQ} ]= -2 \delta^{[M}_{[P}
  R^{a_1 ...a_5  N]}{}_{Q]}+ \delta_{PQ}^{MN} R^{a_2 \ldots a_5, a_1}  \; .
\eqno(2.1.37)$$
The coefficients can be chosen as above and this fixes the normalisations of all the generators.
\par
The action of the Cartan involution was given in  equation (2.1.7), and so
$$
I_c(K^a{}_b)=-K^b{}_a, \ \ I_c(R^M{}_N)=-R^N{}_M, $$
$$ I_c(R^{aMN})= -R_{aMN} ,\ \ I_c(R^{a_1a_2}{}_M )= + R_{a_1a_2}{}^M , {\rm etc} \; .
\eqno(2.1.38)$$
As a result the commutators of the
negative level generators with themselves can be found from those above using the Cartan involution. For example
$$ [R_{aMN}, R_{bPQ}] = \varepsilon_{MNPQR}R_{ab}{}^R \; ,
\eqno(2.1.39)$$
$$ [R_{aMN}, R_{b_1b_2}{}^P] = \delta^P_{[M}R_{ab_1b_2N]} \; ,\ \
{\rm etc}
\eqno(2.1.40)$$
\par
The action of SL(7) on the negative level generators is given by
$$ [K^a{}_b, R_{cMN}] = -\delta^a_cR_{bMN}  \; ,
\eqno(2.1.41)$$
$$ [K^a{}_b, R_{cd}{}^M] = -2\delta^a_{[c}R_{|b|d]}{}^M \; ,
\eqno(2.1.42)$$
$$ [K^a{}_b, R_{a_1a_2a_3M}] = -3\delta_{[a_1}R_{|b|a_2a_3]M} \;
,\ \
\eqno(2.1.43)$$
$$[K^a{}_b,R_{c_1 \ldots c_4}{}^{PQ}] = -4\delta_{[c_1|}^aR_{b|c_2c_3c_4]}{}^{PQ} \; ,
\eqno(2.1.44) $$
$$[K^a{}_b,R_{c_1\ldots c_5M}{}^N]= -5\delta_{[c_1|}^aR_{b| c_2 \ldots c_5] M}{}^N \; ,
\eqno(2.1.45) $$
$$ [K^a{}_b, R_{c_1\ldots c_4,d}]= -4\delta_{[c_1|}^aR_{b|c_2c_3c_4],d} + 4\delta_d^aR_{b[c_1c_2c_3,c_4]} \; .
\eqno(2.1.46) $$

\noindent
While the action of  SL(5) is
$$ [R^M{}_N, R_{aPQ}] = -2\delta^M_{[P}R_{a|N|Q]} + {2\over5}\delta^M_NR_{aPQ} \; ,
\eqno(2.1.47)$$
$$ [R^M{}_N, R_{a_1a_2}{}^P] = \delta^P_NR_{a_1a_2}{}^M - {1\over5}\delta^M_NR_{a_1a_2}{}^P \; ,
\eqno(2.1.48)$$
$$ [R^M{}_N, R_{a_1a_2a_3P}] = -\delta^M_PR_{a_1a_2a_3N} + {1\over 5}\delta^M_NR_{a_1a_2a_3P}\; ,
\eqno(2.1.49)$$
$$ [R^M{}_N, R_{a_1a_2a_3P}] =-\delta_P^MR_{a_1a_2a_3N}+{1\over5}\delta^M_NR_{a_1a_2a_3P} \; ,
\eqno(2.1.50) $$
$$ [R^M{}_N, R_{a_1\ldots a_4}{}^{PQ}] = 2\delta_{N}^{[P|}R_{a_1\ldots a_4}{}^{N|Q]} - {2\over 5}\delta^M_NR_{a_1\ldots a_4}{}^{PQ} \; ,
\eqno(2.1.51)  $$
$$
[R^M{}_N,R_{a_1\ldots a_5P}{}^Q] = -\delta_P^MR_{a_1\ldots a_5N}{}^Q + \delta_N^QR_{a_1\ldots a_5P}{}^M \; ,
\eqno(2.1.52)  $$
$$[R^M{}_N, R_{a_1\ldots a_4,b}] = 0 \; .
\eqno(2.1.53) $$
\par
We now consider the commutators of positive with negative commutators. The generators are all normalised and we can use the Jacobi identities to determine all of these commutators. For example by taking the Jacobi identity involving  two  level one generators and one level -1 generator we find that the following two commutators
$$
[R^{aMN}, R_{bPQ}] = 4\delta^a_b\delta^{[M}_{[P}R^{N]}{}_{Q]} + \delta^{MN}_{PQ} (2 K^a{}_b - {2\over 5} \delta^a_b\sum _cK^c{}_c )  \; ,
\eqno(2.1.54)$$
$$
[R^{a_1a_2}{}_M, R_{a PQ}]= -\varepsilon _{MPQRS}\delta ^{[a_1 }_b R^{a_2] RS}  \; ,
\eqno(2.1.55)$$
The best method is to take the coefficients in these  two equations to be  arbitrary  and then applying the Jacobi identity.
Applying the Cartan involution to equation (2.1.55) we find that
$$ [R^{aMN}, R_{a_1a_2}{}^P] = \varepsilon^{MNPQR}\delta^a_{[a_1}R_{a_2]QR} \; .
\eqno(2.1.56)$$
\par
Similarly using the relevant  the Jacobi identities  we find that
$$ [R^{a_1a_2}{}_M, R_{b_1b_2}{}^N] = 2\delta^{a_1a_2}_{b_1b_2}R^N{}_M -4 \delta^N_M\delta^{[a_1}_{[b_1}K^{a_2]}{}_{b_2]} + {4\over 5} \delta^N_M\delta^{a_1a_2}_{b_1b_2}\sum _e K^e{}_e \; ,
\eqno(2.1.57)$$
While the commutators involving the level $\pm 3$ generators are given by
$$ [R^{a_1a_2a_3M}, R_{bRS}] = 12\delta_b^{[a_1 |}\delta^M_{[R}R^{| a_2a_3]}{}_{S]} \; ,
\eqno(2.1.58)$$
$$ [R_{a_1a_2a_3M}, R^{bRS}] = 12\delta^b_{[a_1 |}\delta_M^{[R}R_{| a_2a_3]}{}^{S]} \; ,
\eqno(2.1.59)$$
$$ [R^{a_1a_2a_3M}, R_{b_1b_2}{}^N] = 12 \delta_{b_1b_2}^{[a_1a_2}R^{a_3]MN} \; ,
\eqno(2.1.60)$$
$$ [ R_{a_1a_2a_3N},R^{b_1b_2}{}_M] = 12\delta^{b_1b_2}_{[a_1a_2}R_{a_3]MN \; , }
\eqno(2.1.61)$$
$$ [R^{a_1a_2a_3M}, R_{b_1b_2b_3N}] = 24\delta^{a_1a_2a_3}_{b_1b_2b_3}R^M{}_N +72\delta^M_N\delta^{[a_1a_2}_{[b_1b_2}K^{a_3]}{}_{b_3]} -{72\over 5} \delta^{a_1a_2a_3}_{b_1b_2b_3}\sum_e K^e{}_e  \; .
\eqno(2.1.62)$$
\par
The commutators involving  level $\pm 4$ generators are given by
$$
[R^{a_1\ldots a_4}{}_{MN},R_{bPQ}] = -2 \varepsilon_{MNPQR}\delta^{[a_1}_bR^{a_2a_3a_4]R} \; ,
 \eqno(2.1.63)  $$
$$
[R_{a_1\ldots a_4}{}^{MN},R^{bPQ}] = -2 \varepsilon^{MNPQR}\delta_{[a_1}^bR_{a_2a_3a_4]R} \; ,
\eqno(2.1.64)  $$
$$
[R^{a_1\ldots a_4}{}_{MN},R_{b_1b_2}{}^P] = 24 \delta^P_M\delta^{[a_1a_2}_{b_1b_2}R^{a_3a_4]}{}_N \; ,
\eqno(2.1.65)  $$
$$
[R_{a_1\ldots a_4}{}^{MN},R^{b_1b_2}{}_P] = 24 \delta^M_P\delta_{[a_1a_2}^{b_1b_2}R_{a_3a_4]}{}^N \; ,
\eqno(2.1.66) $$
$$
[R^{a_1\ldots a_4}{}_{MN},R_{b_1b_2b_3P}] = -24\varepsilon_{MNPQR}\delta^{[a_1a_2a_3}_{b_1b_2b_3}R^{a_4]PQ} \; ,
\eqno(2.1.67)  $$
$$
[R_{a_1\ldots a_4}{}^{MN},R^{b_1b_2b_3P}] = -24\varepsilon^{MNPQR}\delta_{[a_1a_2a_3}^{b_1b_2b_3}R_{a_4]PQ} \; ,
\eqno(2.1.68)  $$
$$[R^{a_1\ldots a_4}{}_{MN}, R_{b_1\ldots b_4}{}^{PQ}] = 96 \delta^{a_1\ldots a_4}_{b_1\ldots b_4}\delta^{[P}_{[M}R^{Q]}{}_{N]}$$
$$ - 192\delta^{PQ}_{MN}\delta^{[a_1a_2a_3}_{[b_1b_2b_3}K^{a_4]}{}_{b_4]}+ {192\over 5}\delta^{PQ}_{MN}\delta^{a_1\ldots a_4}_{b_1\ldots b_4} K^e{}_e \; . \eqno(2.1.69)  $$
\par
Finally, the commutators  involving level $\pm 5$ generators are given by
$$
[R^{a_1\ldots a_5M}{}_N, R_{bPQ}]= 20\delta^M_{[P|}\delta^{[a_1}_bR^{a_2\ldots a_5]}{}_{N|Q]} - 4 \delta^M_N\delta^{[a_1}_bR^{a_2\ldots a_5]}{}_{PQ} \; ,
\eqno(2.1.70)  $$
$$
[R_{a_1\ldots a_5M}{}^N, R^{bPQ}]= 20\delta_M^{[P|}\delta_{[a_1}^bR_{a_2\ldots a_5]}{}^{N|Q]} - 4 \delta_M^N\delta_{[a_1}^bR_{a_2\ldots a_5]}{}^{PQ} \; ,
\eqno(2.1.71)  $$
$$
[R^{a_1\ldots a_5M}{}_N, R_{b_1b_2}{}^P] = 20\delta^P_N\delta^{[a_1a_2}_{b_1b_2}R^{a_3a_4a_4]M} - 4 \delta^M_N\delta^{[a_1a_2}_{b_1b_2}R^{a_3a_4a_5]P} \; , \eqno(2.1.72) $$
$$
[R_{a_1\ldots a_5M}{}^N, R^{b_1b_2}{}_P] = 20\delta_P^N\delta_{[a_1a_2}^{b_1b_2}R_{a_3a_4a_4]M} - 4 \delta_M^N\delta_{[a_1a_2}^{b_1b_2}R_{a_3a_4a_5]P} \; , \eqno(2.1.73)  $$
$$
[R^{a_1\ldots a_5M}{}_N, R_{b_1b_2b_3P}] = 240\delta^M_P\delta^{[a_1a_2a_3}_{b_1b_2b_3}R^{a_4a_5]}{}_N - 48\delta^M_N\delta^{[a_1a_2a_3}_{b_1b_2b_3}R^{a_4a_5]}{}_{P} \; ,
\eqno(2.1.74)  $$
$$
[R_{a_1\ldots a_5M}{}^N, R^{b_1b_2b_3P}] = 240\delta_M^P\delta_{[a_1a_2a_3}^{b_1b_2b_3}R_{a_4a_5]}{}^N - 48\delta_M^N\delta_{[a_1a_2a_3}^{b_1b_2b_3}R_{a_4a_5]}{}^{P} \; ,
\eqno(2.1.75)  $$
$$
[R^{a_1\ldots a_5M}{}_N,R_{b_1\ldots b_4}{}^{PQ}] = 480\delta^{[P|}_N\delta^{[a_1\ldots a_4}_{b_1\ldots b_4}R^{a_5]M|Q]} - 96 \delta^M_N\delta^{[a_1\ldots a_4}_{b_1\ldots b_4}R^{a_5]PQ} \; ,
\eqno(2.1.76)  $$
$$
[R_{a_1\ldots a_5M}{}^N,R^{b_1\ldots b_4}{}_{PQ}] = 480\delta_{[P|}^N\delta_{[a_1\ldots a_4}^{b_1\ldots b_4}R_{a_5]M|Q]} - 96 \delta_M^N\delta_{[a_1\ldots a_4}^{b_1\ldots b_4}R_{a_5]PQ} \; ,
\eqno(2.1.77)  $$
$$
[R^{a_1\ldots a_5M}{}_N,R_{b_1\ldots b_5Q}{}^P] = -480\delta^{a_1\ldots a_5}_{b_1\ldots b_5}(\delta^M_QR^P{}_N - \delta^P_NR^M{}_Q) $$
$$+ 480(5\delta^P_N\delta^M_Q - \delta^P_Q\delta^M_N)\delta^{[a_1\ldots a_4}_{[b_1\ldots b_4}K^{a_5]}{}_{b_5]} -96(5\delta^P_N\delta^M_Q - \delta^P_Q\delta^M_N)\delta^{a_1\ldots a_5}_{b_1\ldots b_5}K^e{}_e  \; ,
\eqno(2.1.78)  $$
$$
[R^{a_1\ldots a_4,b},R_{cMN}] = -{8\over 5} \delta^b_cR^{a_1\ldots a_4}{}_{MN} +{8\over 5} \delta^{[a_1}_cR^{a_2a_3a_4]b}{}_{MN} \; ,
\eqno(2.1.79)  $$
$$
[R_{a_1\ldots a_4,b},R^{cMN}] = -{8\over 5} \delta_b^cR_{a_1\ldots a_4}{}^{MN} +{8\over 5} \delta_{[a_1}^cR_{a_2a_3a_4]b}{}^{MN} \; ,
\eqno(2.1.80)  $$
$$
[R^{a_1\ldots a_4,b},R_{c_1c_2}{}^M] = -{24\over 5}\delta^{[a_1|b}_{c_1c_2}R^{|a_2a_3a_4]M} -{24\over 5} \delta^{[a_1a_2}_{b_1b_2}R^{a_3a_4]bM} \; ,
\eqno(2.1.81)  $$
$$
[R_{a_1\ldots a_4,b},R^{c_1c_2}{}_M] = -{24\over 5}\delta_{[a_1|b}^{c_1c_2}R_{|a_2a_3a_4]M} -{24\over 5} \delta_{[a_1a_2}^{b_1b_2}R_{a_3a_4]bM} \; ,
\eqno(2.1.82)  $$
$$
[R^{a_1\ldots a_4,b},R_{c_1c_2c_3M}] = -{288\over 5} \delta^{[a_1a_2|b}_{c_1c_2c_3}R^{|a_3a_4]}{}_M + {288\over 5}\delta^{[a_1a_2a_3}_{c_1c_2c_3}R^{a_4]b}{}_M \; ,  \eqno(2.1.83) $$
$$
[R_{a_1\ldots a_4,b},R^{c_1c_2c_3M}] = -{288\over 5} \delta_{[a_1a_2|b}^{c_1c_2c_3}R_{|a_3a_4]}{}^M + {288\over 5}\delta_{[a_1a_2a_3}^{c_1c_2c_3}R_{a_4]b}{}^M \; ,  \eqno(2.1.84) $$
$$
[R^{a_1\ldots a_4,b},R_{c_1\ldots c_4}{}^{MN}] = -{192\over 5} \delta^{[a_1a_2a_3|b}_{c_1\ldots c_4}R^{a_4]MN} -{192\over 5} \delta^{a_1\ldots a_4}_{c_1\ldots c_4}R^{bMN} \; ,
\eqno(2.1.85)  $$
$$
[R_{a_1\ldots a_4,b},R^{c_1\ldots c_4}{}_{MN}] = -{192\over 5} \delta_{[a_1a_2a_3|b}^{c_1\ldots c_4}R_{a_4]MN} +-{192\over 5} \delta_{a_1\ldots a_4}^{c_1\ldots c_4}R_{bMN} \; ,
\eqno(2.1.86) $$
$$
[R^{a_1\ldots a_4,b},R_{c_1\ldots c_4,d}] = {384\over 5}(\delta^{a_1\ldots a_4}_{c_1\ldots c_4}K^b{}_d + \delta^{[a_1a_2a_3|b}_{c_1\ldots c_4}K^{|a_4]}{}_d + \delta^{a_1\ldots a_4}_{[c_1c_2c_3|d}K^{b}{}_{|c_4]} $$
$$ + 5\delta^b_d\delta^{[a_1a_2a_3}_{[c_1c_2c_3}K^{a_4]}{}_{c_4]} - 4 \delta^{[a_1a_2a_3|b}_{[c_1c_2c_3|d}K^{|a_4]}{}_{|c_4]}) - {384\over5}(\delta^{a_1\ldots a_4}_{c_1\ldots c_4}\delta^b_d + \delta^{[a_1a_2a_3|b}_{c_1\ldots c_4}\delta^{|a_4]}_d) K^e{}_e \; .
\eqno(2.1.87)  $$

Finally, we find the commutators involving the level 6 generators. We begin with the commutators of level 5 with level 1
$$ [R^{aMN},R^{b_1\ldots b_5 P}{}_{Q}] = \; 2 \varepsilon^{MNPRS} R^{ab_1\ldots b_5}{}_{RS,Q} + 8 \delta^{[N}_QR^{a b_1\ldots b_5 (|P|M])} $$
$$ +20\delta^{[N}_{Q} R^{b_1\ldots b_5,a |P |M]} + 4\delta^{P}_{Q} R^{b_1\ldots b_5,a MN} \; , \eqno(2.1.88)$$
$$[R^{a MN}, R^{b_1\ldots b_4,c}] = 8 R^{b_1\ldots b_4a,c MN} -  R^{ac[b_1b_2 b_3, b_4] MN} \; , \eqno(2.1.89) $$
and then the commutators of the level 6 generators with level -1 are
$$[R_{aMN}, R^{b_1\ldots b_6}{}_{PQ,R}] = \varepsilon_{MNRS[P} \delta^{[b_1}_aR^{b_2\ldots b_6]S}{}_{Q]}+   \varepsilon_{MNPQS} \delta^{[b_1}_aR^{b_2\ldots b_6]S}{}_{R} \; \eqno(2.1.90) $$

$$ [R_{aMN}, R^{b_1\ldots b_6 (PQ)}] = -3 \delta^{(Q}_{[N}\delta^{[b_1}_aR^{b_2\ldots  b_6] P)}{}_{M]} \; , \eqno(2.1.91)$$

$$ [R_{aMN},R^{b_1\ldots b_5,c PQ}] = - {1\over 3  } \delta^{[Q}_{[R} \delta^{[b_1}_aR^{b_2\ldots b_5],c P]}{}_{M]} $$ 
$$-{1\over 3 } \delta^{[Q}_{[R} \delta^{c}_aR^{b_1\ldots b_5 P]}{}_{M]} + \delta^{MN}_{PQ} \delta^{[b_1}_aR^{b_2\ldots b_5],c} \; . \eqno(2.1.91) $$

\medskip
{\bf 2.2 The commutators of $E_{11}$ with the vector  representation}
\medskip
\noindent
The elements of the vector ($l_1$) representation  when decomposed into representations of
$SL(7)\otimes SL(5) $ are given by
$$ P_a; \ \ Z^{MN} ; \ \ Z^a{}_M; \ \ Z^{a_1a_2M}; \ \ Z^{a_1a_2a_3}{}_{MN} ; \ \ Z^{a_1a_2a_3,b}, \ \ Z^{a_1\ldots a_4} , \ \ Z^{a_1\ldots a_4 M}{}_{N} ;
$$
$$
 \ \ Z^{a_1\ldots a_5 MN}, \ \ Z^{a_1\ldots a_5 (MN)} , \ \ Z^{a_1\ldots a_5}{}_{MN,P}, \ \ Z^{a_1\ldots a_4,bMN}, \ldots   \eqno(2.2.1)$$
These belong to irreducible representations of  $SL(7)\otimes SL(5) $ and our index conventions are as given earlier for the $E_{11}$ generators. We now regard these as generators whose commutator is given by equation (2.1.5). They have a level that is the number of up minus down indices  plus one. The generators in the vector representation at low levels and the coordinates they lead to in the non-linear realisation we given in references [11,12,13].

\par
As they belong to representation of SL(7), their  commutators of the generators  of  SL(7) are given by
$$ [K^a{}_b, P_c] = -\delta^a_cP_b + {1\over2}\delta^a_bP_c  \; ,
\eqno(2.2.2)$$
$$ [K^a{}_b, Z^{MN}] = {1\over2}\delta^a_bZ^{MN} \; ,
\eqno(2.2.3)$$
$$ [K^a{}_b, Z^c{}_M] = \delta^c_bZ^a{}_M + {1\over 2}\delta^a_bZ^c{}_M  \; ,
\eqno(2.2.4)$$
$$ [K^a{}_b, Z^{a_1a_2M}] = 2\delta_b^{[a_1}Z^{|a|a_2]M} + {1\over2}\delta^a_bZ^{a_1a_2M} \; ,
\eqno(2.2.5)$$
$$
[K^a{}_b, Z^{b_1b_2b_3}{}_{PQ}] = 3 \delta^{[b_1|}_bZ^{a|b_2b_3]}{}_{PQ} + {1\over 2} \delta^a{}_bZ^{b_1b_2b_3}{}_{PQ}  \; ,  \eqno(2.2.6) $$
$$
[K^a{}_b, Z^{c_1\ldots c_4M}{}_N] = 4\delta^{[c_1|}_bZ^{a|c_2c_3c_4]M}{}_N + {1\over 2}\delta^a{}_bZ^{c_1\ldots c_4M}{}_{N}  \; ,  \eqno(2.2.7) $$
$$
[K^a{}_b, Z^{c_1c_2c_3,d}] = 3 \delta^{[c_1|}_b Z^{a|c_2c_3],d} + \delta^d_bZ^{c_1c_2c_3,a} +  {3\over2}\delta^a_bZ^{d[c_1c_2,c_3]}  \; ,  \eqno(2.2.7) $$
$$
[K^a{}_b,Z^{c_1\ldots c_4}] = 4\delta^{[c_1|}_bZ^{a|c_2c_3]} + {1\over 2} \delta^a_bZ^{c_1\ldots c_4}  \; .  \eqno(2.2.8)  $$
\par
While with  the generators of SL(5) we have
$$ [R^M{}_N, P_a] = 0  \; ,
\eqno(2.2.9)$$
$$ [R^M{}_N, Z^{PQ}] = 2\delta_N^{[P}Z^{|M|Q]} - {2\over5}\delta^M_NZ^{PQ}  \; ,
\eqno(2.2.10)$$
$$ [R^M{}_N, Z^a{}_P] = -\delta^M_PZ^a{}_N + {1\over 5}\delta^M_NZ^a{}_P  \; ,
\eqno(2.2.11)$$
$$ [R^M{}_N, Z^{a_1a_2P}] = \delta^P_NZ^{a_1a_2M} -{1\over 5}\delta^M_NZ^{a_1a_2P}  \; ,
\eqno(2.2.12)$$
$$
[R^M{}_N,Z^{a_1a_2a_3}{}_{PQ}] = -2\delta^M_{[P|}Z^{a_1a_2a_3}{}_{N|Q]} + {2\over5}\delta^M_NZ^{a_1a_2a_3}{}_{PQ}  \; ,  \eqno(2.2.13)  $$
$$
[R^M{}_N, Z^{c_1\ldots c_4P}{}_Q ] = \delta^P_NZ^{c_1\ldots c_4M}{}_Q - \delta^M_QZ^{c_1\ldots c_4P}{}_N   \; ,  \eqno(2.2.14) $$
$$
[R^M{}_N, Z^{c_1c_2c_3,d}] = 0  \; ,  \eqno(2.2.15) $$
$$
[R^M{}_N,Z^{c_1\ldots c_4}] = 0  \; ,  \eqno(2.2.16)  $$
\par
The commutators of the $E_{11}$ generators with those of the vector representation can be found using similar arguments to those used to find the commutators of $E_{11}$. The commutators must preserve the level and their $SL(7)\otimes SL(5) $ character must be the same on both sides of the commutator. The repeated commutators of the level one $E_{11}$ generator are  used to define the normalisation of the  $l_1$ generators as follows
$$ [R^{aMN}, P_b] = \delta^a_bZ^{MN}  \; ,
\eqno(2.2.17)$$
$$ [R^{aMN}, Z^{PQ}] = -\varepsilon^{MNPQR}Z^a{}_R  \; ,
\eqno(2.2.18)$$
$$ [R^{aMN}, Z^b{}_P] = 2\delta_P^{[M}Z^{abN]}  \; ,
\eqno(2.2.19)$$
$$ [ R^{aMN}, Z^{b_1b_2 P} ] = \varepsilon^{MNPRS} Z^{ab_1b_2 }{}_{RS} \; ,
\eqno(2.2.20)$$
$$ [R^{aMN},Z^{b_1b_2 b_3}{}_{ PQ} ] =\delta _{PQ}^{MN} (Z^{b_1b_2b_3 , a}
+ Z^{b_1b_2b_3  a} )+ \delta_{[P}^{[M | } Z^{ab_1b_2b_3 |N]} {}_{Q]}  \; .
\eqno(2.2.21)$$
\par
Using the $E_{11}$ commutators and the Jacobi identities we can find the commutators involving the positive $E_{11}$ generators  to be as follows
$$ [R^{a_1a_2}{}_P, P_b] = 2\delta_b^{[a_1}Z^{a_2]}{}_P   \; ,
\eqno(2.2.22)$$
$$[R^{a_1a_2}{}_P, Z^{RS}] = 2\delta_P^{[R}Z^{a_1a_2 S]}  \; ,
\eqno(2.2.23)$$
$$[R^{a_1a_2}{}_P, Z^{b}{}_S] = 2 Z^{a_1a_2 b}{}_{PS}  \; ,
\eqno(2.2.24)$$
$$[R^{a_1a_2}{}_P, Z^{b_1 b_2 R}] = -2\delta_P^R ( Z^{a_1a_2 b_1b_2} + Z^{b_1b_2  [a_1 , a_2]} ) - {1\over 2} Z^{a_1a_2 b_1b_2}{}^R{}_{P}  \; ,
\eqno(2.2.25)$$
For example to find the first relation we use the Jacobi identity $[[R^{a_1MN},R^{a_2PQ}], P_b]+\ldots =0$.
\par
Similarly the commutators involving the level three $E_{11}$ generators are given by
$$ [R^{a_1a_2a_3M}, P_b] = -6 \delta_b^{[a_1}Z^{a_2a_3]M}  \; ,
\eqno(2.2.26)$$
$$ [R^{a_1a_2a_3M}, Z^{PQ}] = -2\varepsilon^{MPQRS} Z^{a_1a_2a_3} {}_{RS} \; ,
\eqno(2.2.27)$$
$$ [R^{a_1a_2a_3M}, Z^b{}_Q] =  Z^{a_1a_2a_3b} {}^M{}_Q+6 \delta^M_Q Z^{b [a_1a_2 ,a_3]} - 6 \delta^M_Q Z^{a_1a_2a_3b} \; ,
\eqno(2.2.28)$$
while those involving the  level four generators by
$$ [R^{a_1\ldots a_4}{}_{M N}, P_b] = 8\delta_b^{[a_1 }Z^{a_2a_3a_4 ]} {}_{MN}  \; ,
\eqno(2.2.29)
$$
$$
[R^{a_1 \ldots a_4}{}_{M N}, Z^{PQ}]=8 \delta_{MN}^{PQ}Z^{a_1\ldots a_4}
+2 \delta^{[P |} _{[M |}Z^{a_1\ldots a_4 |Q ]}{}_{|N]}  \; ,
\eqno(2.2.30)$$
and the commutators with  level five generators are given by
$$
[ R^{a_1\ldots a_5 M}{}_{N}, P_c ]= 5\delta^{ [ a_1 }_c Z^{a_2\ldots a_5] M}{}_{N} \; ,
\eqno(2.2.31)$$
$$
[ R^{a_1\ldots a_4 , b} , P_c ] = 8 \delta^{ [ a_1}_c Z^{a_2a_3a_4 ] , b} +8 (\delta^{ [ a_1}_c Z^{a_2a_3a_4 ]  b} -\delta_c^b Z^{a_1 \ldots a_4}) \; .
\eqno(2.2.32)
$$
\par
We now turn to consider the commutators of the negative level $E_{11}$ generators with those of the vector representation. As the vector representation is a lowest weight representation we have by definition the commutator
$$ [R_{aMN}, P_b] = 0  \; ,
\eqno(2.2.33)$$
By considering the Jacobi identity $[R_{aMN},[R^{bPQ}, P_c]]+\ldots =0$
we find the commutator
$$ [R_{aMN}, Z^{PQ}] = 2\delta^{PQ}_{MN}P_a  \; ,
\eqno(2.2.34)$$
Using similar arguments we find that  the commutators of level minus one $E_{11}$ generators   with those of the vector representation
take the form
$$ [R_{aMN}, Z^b{}_P] = -{1\over2}\delta^a_b\varepsilon_{MNPQR}Z^{QR}  \; ,
\eqno(2.2.35)$$
$$ [R_{aMN}, Z^{b_1b_2P}] = -4\delta^P_{[M}\delta_a^{[b_1}Z^{b_2]}{}_{N]} \; ,
\eqno(2.2.36)$$
$$ [R_{aMN}, Z^{b_1b_2b_3}{}_{PQ}] = {3\over 2} \varepsilon_{MNPQ R}\delta^{[b_1}_a Z^{b_2b_3]} {}^R \; ,
\eqno(2.2.37)$$
$$ [R_{aMN}, Z^{c_1c_2c_3, b} ]= {3\over 2} (\delta_a^{[c_1} Z^{c_2c_3]b}{}_{MN} +\delta_a^b Z^{c_1c_2c_3}{}_{MN}) \; ,
\eqno(2.2.38)$$
$$ [R_{aMN}, Z^{c_1 \ldots c_4 }]= -{2\over 5} \delta_a^{[c_1} Z^{c_2c_3c_4 ]}{}_{MN} \; ,
\eqno(2.2.39)$$
$$ [R_{aMN}, Z^{c_1\ldots c_4}{}^Q{}_S ]= -32(\delta_{[M |}^Q \delta_a^{[c_1} Z^{c_2c_3c_4 ]}{}_{ |N]S}
+{1\over 5} \delta_{S}^Q \delta_a^{[c_1} Z^{c_2c_3c_4 ]}{}_{MN} ) \; ,
\eqno(2.2.40)$$
\par
Using equation (2.1.40) and the corresponding Jacobi identity we find the commutators of the level minus two generators to be given by
$$ [R_{a_1a_2}{}^P, P_b] = 0= [R_{a_1a_2}{}^P, Z^{RS} ] \; ,
\eqno(2.2.41)$$
$$ [R_{a_1a_2}{}^P, Z^b{}_R] = 2\delta^P_R\delta^b_{[a_1}P_{a_2]}  \; ,
\eqno(2.2.42)$$
$$ [R_{a_1a_2}{}^P, Z^{b_1b_2 R}] = -2\delta^{b_1b_2}_{a_1a_2}Z^{PR}  \; ,
\eqno(2.2.43)$$
$$ [R_{a_1a_2}{}^P, Z^{b_1b_2 b_3}{}_{ RS}] = -6\delta^{[ b_1b_2 |}_{a_1a_2}
\delta_{[R |}^P Z^{ |b_3 ]} {}_{ |S]}  \; ,
\eqno(2.2.44)$$
$$
[ R_{a_1 a_2}{}^L , Z^{b_1\ldots b_4}  ]= {6\over 5} \delta_{a_1 a_2}^{[b_1b_2} Z^{b_3b_4] L}  \; ,
\eqno(2.2.45)$$
$$
[ R_{a_1 a_2}{}^L , Z^{b_1 b_2b_3 , c}  ]= -3 (\delta_{a_1 a_2}^{[b_1b_2} Z^{b_3 ] c L} +  \delta_{a_1 a_2}^{c [b_1} Z^{b_2 b_3 ]  L}) \; ,
\eqno(2.2.46)$$
$$
[ R_{a_1 a_2}{}^L , Z^{b_1 \ldots b_4}{}^R{}_S  ]= 24  \delta_{a_1 a_2}^{[b_1b_2 |} ( \delta^S_L Z^{ | b_3 b_4 ] R}
-{1\over 5}  \delta_S^R Z^{ | b_3 b_4 ]  L}) \; ,
\eqno(2.2.47)$$
while  the commutators  involving  level minus three $E_{11}$ generators are given by by
$$ [R_{a_1a_2a_3}{}_P, P_b] = 0= [R_{a_1a_2a_3}{}_P, Z^{RS} ]= [R_{a_1a_2a_3}{}_P, Z^{b}{}_S ] \; ,
\eqno(2.2.48)$$
$$ [R_{a_1a_2a_3M}, Z^{b_1b_2N}] = -12 \delta^N_M\delta^{b_1b_2}_{[a_1a_2}P_{a_3]}  \; ,
\eqno(2.2.49)$$

$$ [R_{a_1a_2a_3M}, Z^{b_1b_2b_3}{}_{NP}] = -3 \varepsilon_{MNPQR}\delta^{b_1b_2b_3}_{a_1a_2a_3}Z^{QR}  \; ,
\eqno(2.2.50)$$
$$
[R_{a_1a_2a_3M}, Z^{b_1 \ldots b_4} ]= -{3!\cdot 3!\over 5}\delta_{a_1a_2a_3 }^{[c_1c_2c_3} Z^{c_4]}{}_M  \; ,
\eqno(2.2.51)$$
$$
[R_{a_1a_2a_3M}, Z^{b_1b_2b_3, c } ]= {9}(\delta_{a_1a_2a_3}^{b_1b_2b_3} Z^c{}_M +\delta_{a_1a_2a_3}^{c [ b_1b_2 }Z^{b_3]}{}_M)  \; ,
\eqno(2.2.52)$$
$$
[R_{a_1a_2a_3M}, Z^{b_1\ldots b_4 Q}{}_{T} ]= 4\cdot 4! (\delta_M^Q \delta_{a_1a_2a_3 }^{[b_1b_2b_3} Z^{b_4]}{}_{T}-{1\over 5}
\delta_T^Q \delta_{a_1a_2a_3 }^{[b_1b_2b_3} Z^{b_4]}{}_{M})  \; .
\eqno(2.2.53)$$
\par
The commutators of the level minus four $E_{11}$ generators with the generators of the vector representation are given by
$$
[ R_{a_1\ldots a_4} {}^{S_1S_2} , Z^{b_1b_2b_3}{}_{L_1L_2}]
=-4! \delta^{S_1S_2} _{L_1L_2} P_{[ a_1}\delta _{a_2a_3a_4]}^{b_1b_2b_3}  \; ,
\eqno(2.2.54)$$
$$
[ R_{a_1\ldots a_4} {}^{S_1S_2} , Z^{b_1\ldots b_4} ]= -{4!\over 5}  \delta_{a_1\ldots a_4}^{b_1\ldots b_4}Z^{S_1S_2}  \; ,
\eqno(2.2.55)$$
$$
[ R_{a_1\ldots a_4} {}^{S_1S_2} , Z^{b_1b_2b_3, c} ]=0  \; ,
\eqno(2.2.56)$$
$$
[ R_{a_1\ldots a_4} {}^{S_1S_2} , Z^{b_1\ldots b_4} {}^{R}{}_{T} ]
=4\cdot 4! \delta_{a_1\ldots a_4}^{b_1\ldots b_4}(\delta_{T}^{[S_1} Z^{S_2]R}+{1\over 5} \delta_{T}^{R}Z^{S_1S_2} ) \; ,
\eqno(2.2.57)$$
and finally with the level  minus five  $E_{11}$ generators are given by
$$
[ R_{a_1\ldots a_5 R }{}^S, Z^{b_1\ldots b_4}]= 0 \; ,
\eqno(2.2.58)$$
$$
[ R_{a_1\ldots a_4,b} , Z^{c_1\ldots c_4} ]= -{48 \over 5} (P_{b}\delta_{a_1\ldots a_4}^{c_1\ldots c_4} +P_{[a_1 |}\delta_{b |a_2a_3 a_4 ]}^{c_1\ldots c_4}) \; ,
\eqno(2.2.59)$$
$$
[ R_{a_1\ldots a_4,b} , Z^{c_1 c_2 c_3,d} ]= 36(P_{[a_1 |}\delta_{b}^{[ c_1}\delta_{ |a_2a_3 a_4 ]}^{c_2 c_3 ]d} +\delta_b^d P_{[a_1 }\delta_{a_2a_3 a_4 ]}^{c_1c_2 c_3})  \; ,
\eqno(2.2.60)$$
$$
[ R_{a_1\ldots a_5R}{}^S, Z^{b_1b_2b_3,c}]= 0 \; ,
\eqno(2.2.61) $$
$$
[R_{a_1\ldots a_5M}{}^N , Z^{b_1\ldots b_4R}{}_S] = -96(\delta^N_M\delta^R_T - 5\delta^N_T\delta^R_M)\delta^{b_1\ldots b_4}_{[a_1\ldots a_4}P_{a_5]}  \; ,
 \eqno(2.2.62)  $$
$$
[ R_{a_1\ldots a_4,b} , Z^{c_1\ldots c_4M}{}_N] = 0  \; .
 \eqno(2.2.63)  $$

\medskip
{\bf 2.3 The Cartan involution invariant algebra,  $I_c(E_{11})$}
\medskip
This algebra plays a crucial role in constructing the invariant dynamics. The algebra $I_c(E_{11})$ at level zero is  just the Cartan involution  invariant suabalgeba of $GL(7) \otimes SL(5)$ which is $SO(7) \otimes SO(5)$. The Cartan involution invariant generators are
$$
J^a{}_b = K^a{}_b - K^b{}_a; \ \ S^M{}_N = R^M{}_N - R^N{}_M;
$$
$$
S^{aMN} = R^{aMN} - R_{aMN}; \ \ S^{a_1a_2}{}_M = R^{a_1a_2}{}_M + R_{a_1a_2}{}^M;
$$
$$ S^{a_1a_2a_3M} = R^{a_1a_2a_3M} - R_{a_1a_2a_3M} \; ; \; \ldots
\eqno(2.3.1)$$
In the following, we will sometimes refer to these generators as 'even' generators.
\par
The generators at level zero obey the algebra of $SO(7) \otimes SO(5)$:
$$[J^a{}_b, J^c{}_d] = \delta^c_bJ^a{}_d - \delta^c_aJ^b{}_d - \delta^a{}_dJ^c{}_b + \delta^b_dJ^c{}_a  \; ,
\eqno(2.3.2)$$
$$[S^M{}_N, S^P{}_Q] = \delta^P_NS^M{}_Q - \delta^P_MS^N{}_Q - \delta^M_QS^P{}_N + \delta^N_QS^P{}_M  \; ,
\eqno(2.3.3)$$
$$
[J^a{}_b, S^M{}_N] = 0  \; .
\eqno(2.3.4)$$
\par
The commutators of the generators of $SO(7) \otimes SO(5)$ with the other generators of $I_c(E_{11})$
are determined by the representations to which the latter belong and are of standard form which is given by
$$
[J^a{}_b, S^{cMN}] = \delta^c_bS^{aMN} - \delta^c_aS^{bMN}  \; ,
\eqno(2.3.5)$$
$$
[J^a{}_b, S^{cd}{}_M] = 2\delta_b^{[c}S^{|a|d]}{}_M - 2 \delta_a^{[c}S^{|b|d]}{}_M \; ,
\eqno(2.3.6)$$
$$
[J^a{}_b, S^{c_1c_2c_3M}] = 3\delta_b^{[c_1}S^{|a|c_2c_3]M} - 3\delta_a^{[c_1}S^{|b|c_2c_3]M}  \; ,
\eqno(2.3.7)$$
$$
[S^M{}_N, S^{aPQ}] = 2\delta_N^{[P}S^{a|M|Q]} - 2\delta_M^{[P}S^{a|N|Q]}  \; ,
\eqno(2.3.8)$$
$$ [S^M{}_N, S^{ab}{}_P] = -\delta^M_PS^{ab}{}_N + \delta^N_PS^{ab}{}_M  \; ,
\eqno(2.3.9)$$
$$
[S^M{}_N, S^{a_1a_2a_3P}] = \delta^P_NS^{a_1a_2a_3M} - \delta^P_MS^{a_1a_2a_3N}  \; ,
\eqno(2.3.10)$$
\par
The commutators of the generators of $I_c(E_{11})$ are easily found by using the commutators of $E_{11}$ given in  section (2.1)  and their definition of equation (2.3.1). We find that
$$
[S^{aMN}, S^{bPQ}] = \varepsilon^{MNPQR}S^{ab}{}_R -4 \delta^a_b\delta^{[M}_{[P}S^{N]}{}_{Q]} - 2\delta^{MN}_{PQ}J^a{}_b  \; ,
\eqno(2.3.11)$$
$$
[S^{aMN}, S^{b_1b_2}{}_P] = \delta_P^{[M}S^{ab_1b_2N]} -\varepsilon^{MNPQR}\delta_a^{[b_1}S^{b_2]QR}  \; ,
\eqno(2.3.12)$$
$$
[S^{aMN}, S^{b_1b_2b_3P}] = 12\delta_a^{[b_1}\delta^P_{[M}S^{b_2b_3]}{}_{N]}+ \varepsilon ^{MNPL_1L_2} S^{a b_1b_2b_3 }{}_{L_1L_2}   \; ,
\eqno(2.3.13)$$
$$
[S^{a_1a_2}{}_M, S^{b_1b_2}{}_N] = S^{a_1a_2b_1b_2}{}_{MN}-{ 2}\delta^{a_1a_2}_{b_1b_2}S^M{}_N - 4\delta^M_N\delta^{[a_1}_{[b_1}J^{a_2]}{}_{b_2]}  \; ,
\eqno(2.3.14)$$
$$
[S^{a_1a_2}{}_M, S^{b_1b_2b_3N} ]= S^{a_1a_2b_1b_2b_3 N}{}_M- 12\delta_{a_1a_2}^{[b_1b_2}S^{b_3]MN}  \; ,
\eqno(2.3.15)$$
$$[S^{a_1a_2a_3M}, S^{b_1b_2b_3N}] = -24\delta^{a_1a_2a_3}_{b_1b_2b_3}S^M{}_N - 72\delta^M_N\delta^{[a_1a_2}_{[b_1b_2}J^{a_3]}{}_{b_3]}  \; .
\eqno(2.3.16)$$


\medskip
{\bf 2.4 The $l_1$ representation with the representation of $I_c(E_{11})$ }
\medskip
In this section we  give the commutators with the $I_c(E_{11})$ generators with those of the $l_1$ representation.  These are easily computed using the commutators of the generators of $E_{11}$ with the vector representation which are given in section (2.2). The commutators of   the generators of $SO(7) \otimes SO(5)$ with the vector representation are given by
$$
[J^a{}_b, P_c] = - \delta^a_cP_b + \delta^b_cP_a  \; ,
\eqno(2.4.1)$$
$$
[J^a{}_b, Z^{MN}] = 0  \; ,
\eqno(2.4.2)$$
$$
[J^a{}_b, Z^c{}_M] = \delta^c_b Z^a{}_M - \delta^c_aZ^b{}_M \; ,
\eqno(2.4.3)$$
$$
[J^a{}_b, Z^{c_1c_2M}] = 2\delta_b^{[c_1}Z^{|a|c_2]M} - 2\delta_a^{[c_1}Z^{|b|c_2]M} \; ,
\eqno(2.4.4)$$
$$
[S^M{}_N, P_a] = 0 \; ,
\eqno(2.4.5)$$
$$
[ S^M{}_N, Z^{PQ}] = 2\delta_N^{[P}Z^{|M|Q]} - 2\delta_M^{[P}Z^{|N|Q]} \; , \ \ \ {\rm etc}\
\eqno(2.4.6)$$
\par
The commutators of the generators of  $I_c(E_{11})$ involving  level $\pm 1$ generators  with the vector representation are given by
$$
[S^{aMN}, P_b] = \delta^a_bZ^{MN} \; ,
\eqno(2.4.7)$$
$$
[S^{aMN}, Z^{PQ}] = - \varepsilon^{MNPQR}Z^a{}_R - 2\delta^{PQ}_{MN}P_a \; ,
\eqno(2.4.8)$$
$$
[S^{aMN}, Z^b{}_P] = 2\delta_P^{[M}Z^{abN]} + {1\over2}\delta^a_b\varepsilon_{MNPQR}Z^{QR} \; ,
\eqno(2.4.9)$$
$$[S^{aMN}, Z^{b_1b_2P}] = \varepsilon^{MNPRS} Z^{ab_1b_2}{}_{RS}+4\delta^P_{[M}\delta_a^{[b_1}Z^{b_2]}{}_{N]} \; ,
\eqno(2.4.10)$$
$$[S^{aMN}, Z^{b_1b_2b_3}{}_{PQ}] = \delta _{PQ}^{MN} (Z^{b_1b_2b_3 , a}
+ Z^{b_1b_2b_3  a} ) $$
$$+ \delta_{[P}^{[M | } Z^{ab_1b_2b_3 |N]} {}_{Q]}
-{3\over 2} \varepsilon_{MNPQR}\delta_a^{[b_1} Z^{b_2b_3]} {}^R  \; ,
\eqno(2.4.11)$$
$$ [S^{aMN}, Z^{c_1c_2c_3, b} ]= -{3\over 2} (\delta_a^{[c_1} Z^{c_2c_3]b}{}_{MN} +\delta_a^b Z^{c_1c_2c_3}{}_{MN}) + \ldots \; ,
\eqno(2.4.12)$$
$$ [S^{aMN}, Z^{c_1\ldots c_4 }]= {2\over 5} \delta_a^{[c_1} Z^{c_2c_3c_4 ]}{}_{MN} + \ldots  \; ,
\eqno(2.4.13)$$
$$ [S^{aMN}, Z^{c_1 \ldots c_4}{}^Q{}_S ]= 8\cdot4(\delta_{[M |}^Q \delta_a^{[c_1} Z^{c_2c_3c_4 ]}{}_{ |N]S}
+{1\over 5} \delta_{S}^Q \delta_a^{[c_1} Z^{c_2c_3c_4 ]}{}_{MN} ) \ldots  \; .
\eqno(2.4.14)$$
The $+\ldots $ in the above equations  indicate the presence of generators  which are of a  higher level than we are considering in this paper.
\par
The commutators of the generators of  $I_c(E_{11})$ subalgebra involving  level $\pm 2$ generators  with those of the  vector representation are given by
$$
[S^{a_1a_2}{}_P, P_b] = 2\delta_b^{[a_1}Z^{a_2]}{}_P  \; ,
\eqno(2.4.15)$$
$$
[S^{a_1a_2}{}_P, Z^{RS}] = 2\delta_P^{[R |}Z^{a_1a_2 |S]}  \; ,
\eqno(2.4.16)$$
$$
[S^{a_1a_2}{}_P, Z^b{}_S] =2Z^{a_1a_2 b}{}_{PS}+ 2 \delta^P_S\delta^b_{[a_1}P_{a_2]}  \; ,
\eqno(2.4.17)$$
$$
[S^{a_1a_2}{}_P, Z^{b_1b_2 R}] = -2\delta_P^R ( Z^{a_1a_2 b_1b_2}+Z^{b_1b_2  [a_1 , a_2]}) -{1\over 2} Z^{a_1a_2 b_1b_2}{}^R
{}_{P}
-2\delta^{b_1b_2}_{a_1a_2}Z^{PR}  \; ,
\eqno(2.4.18)$$
$$ [S_{a_1a_2}{}^P, Z^{b_1b_2 b_3}{}_{ RS}] = -6\delta^{[ b_1b_2 |}_{a_1a_2}
\delta_{[R |}^P Z^{ |b_3 ]} {}_{ |S]}+\ldots
\eqno(2.4.19)$$
$$
[ S^{a_1 a_2}{}_L , Z^{b_1\ldots b_4}  ]= {6\over 5} \delta_{a_1 a_2}^{[b_1b_2} Z^{b_3b_4] L} +\ldots \; ,
\eqno(2.4.20)$$
$$
[ S^{a_1 a_2}{}_L , Z^{b_1 b_2b_3 , c}  ]= -3 (\delta_{a_1 a_2}^{[b_1b_2} Z^{b_3 ] c L} +  \delta_{a_1 a_2}^{c [b_1} Z^{b_2 b_3 ]  L})+\ldots  \; ,
\eqno(2.4.21)$$
$$
[ S^{a_1 a_2}{}_L , Z^{b_1 \ldots b_4}{}^R{}_S  ]= 8\cdot 3  \delta_{a_1 a_2}^{[b_1b_2 |} ( \delta_S^L Z^{ | b_3 b_4 ] R}
-{1\over 5}  \delta_S^R Z^{ | b_3 b_4 ]  L})+\ldots  \; .
\eqno(2.4.22)$$
\par
The commutators of the generators of  $I_c(E_{11})$ subalgebra involving  level $\pm 3$ generators  with those of the  vector representation are given by
$$
[S^{a_1a_2a_3M}, P_b] = - 6\delta_b^{[a_1}Z^{a_2a_3]M}  \; ,
\eqno(2.4.23)$$
$$
[S^{a_1a_2a_3M}, Z^{PQ}] = -2\varepsilon^{MPQRS} Z^{a_1a_2a_3}{}_{RS} \; ,
\eqno(2.4.24)$$
$$
[S^{a_1a_2a_3M}, Z^b{}_N] = Z^{a_1a_2a_3b} {}^M{}_N+6\delta^M_N Z^{b [a_1a_2 ,a_3]}-6 \delta^M_N Z^{a_1a_2a_3b} \; ,
\eqno(2.4.25)$$
$$
[S^{a_1a_2a_3M},Z^{b_1b_2N}] = 12\delta^N_M\delta^{b_1b_2}_{[a_1a_2}P_{a_3]} +\ldots  \; ,
\eqno(2.4.26)$$
$$ [S^{a_1a_2a_3M}, Z^{b_1b_2b_3}{}_{NP}] = 3 \varepsilon_{MNPQR}\delta^{b_1b_2b_3}_{a_1a_2a_3}Z^{QR}+\ldots  \; ,
\eqno(2.4.27)$$
$$
[S^{a_1a_2a_3M}, Z^{b_1\ldots  b_4} ]= {3!\cdot 3!\over 5}\delta_{a_1a_2a_3 }^{[c_1c_2c_3} Z^{c_4]}{}_M +\ldots  \; ,
\eqno(2.4.28)$$
$$
[S^{a_1a_2a_3M}, Z^{b_1b_2b_3, c } ]=- {9}(\delta_{a_1a_2a_3}^{b_1b_2b_3} Z^c{}_M +\delta_{a_1a_2a_3}^{c [ b_1b_2 }Z^{b_3]}{}_M)+\ldots  \; ,
\eqno(2.4.29)$$
$$
[S^{a_1a_2a_3M}, Z^{b_1\ldots b_4 Q}{}_{T} ]= -4\cdot 4! (\delta_M^Q \delta_{a_1a_2a_3 }^{[b_1b_2b_3} Z^{b_4]}{}_{T}-{1\over 5}
\delta_T^Q \delta_{a_1a_2a_3 }^{[b_1b_2b_3} Z^{b_4]}{}_{M})+\ldots  \; .
\eqno(2.4.30)$$
\par
The commutators of the generators of  $I_c(E_{11})$ subalgebra involving  level $\pm 4$ generators  with those of the  vector representation are given by
$$
[S^{a_1\ldots a_4}{}_{MN},P_b] = 8\delta^{[a_1}_bZ^{a_2 a_3a_4 ]}{}_{MN}  \; ,
\eqno(2.4.31)$$
$$
[S^{a_1\ldots a_4}{}_{M N}, Z^{PQ}]= 8 \delta_{MN}^{PQ}Z^{a_1\ldots a_4}
+2 \delta^{[P |} _{[M |}Z^{a_1\ldots a_4 |Q ]}{}_{|N]} +\ldots    \; ,
\eqno(2.4.32)$$
$$
[ S^{a_1\ldots a_4} {}_{S_1S_2} , Z^{b_1b_2b_3}{}_{L_1L_2}]
=-4! \delta^{S_1S_2} _{L_1L_2} P_{[ a_1}\delta _{a_2a_3a_4]}^{b_1b_2b_3}+\ldots \; ,
\eqno(2.4.33)$$
$$
[ S^{a_1\ldots a_4} {}_{S_1S_2} , Z^{b_1\ldots b_4} ]=- {4!\over 5}  \delta_{a_1\ldots a_4}^{b_1\ldots b_4}Z^{S_1S_2} +\ldots \; ,
\eqno(2.4.34)$$
$$
[ S^{a_1\ldots a_4} {}_{S_1S_2} , Z^{b_1b_2b_3, c} ]=0+\ldots \; ,
\eqno(2.4.35)$$
$$
[ S^{a_1\ldots a_4} {}_{S_1S_2} , Z^{b_1\ldots b_4} {}^{R}{}_{T} ]
=4\cdot 4! \delta_{a_1\ldots a_4}^{b_1\ldots b_4}(\delta_{T}^{[S_1} Z^{S_2]R}+{1\over 5} \delta_{T}^{R}Z^{S_1S_2} )+\ldots \; .
\eqno(2.4.36)$$
\noindent
Finally The commutators of the generators of  $I_c(E_{11})$ subalgebra involving  level $\pm 5$ generators  with those of the  vector representation are given by
$$
[ S^{a_1\ldots a_4 , b} , P_c ] = 8 \delta^{ [ a_1}_c Z^{a_2a_3a_4 ] , b}
+8 (\delta^{ [ a_1}_c Z^{a_2a_3a_4 ]  b} -\delta_c^b Z^{a_1\ldots  a_4}) \; ,
\eqno(2.4.37)
$$
$$
[ S^{a_1\ldots a_5 M}{}_{N}, P_c ]= 5\delta^{ [ a_1 }_c Z^{a_2\ldots a_5] M}{}_{N} \; ,
\eqno(2.4.38)$$
$$
[ S^{a_1\ldots a_5 R}{}_S, Z^{b_1\ldots b_4}]= 0+\ldots  \; ,
\eqno(2.4.39)$$
$$
[ S^{a_1\ldots a_4,b} , Z^{b_1\ldots b_4} ]= {48\over 5} (P_{b}\delta_{a_1\ldots a_4}^{b_1\ldots b_4} +P_{[a_1 |}\delta_{b |a_2a_3 a_4 ]}^{b_1\ldots b_4})+\ldots  \; ,
\eqno(2.4.40)$$
$$
[ S^{a_1\ldots a_4,b} , Z^{c_1 c_2 c_3,d} ]=- 3!\cdot 3!(P_{[a_1 |}\delta_{b}^{[ c_1}\delta_{ |a_2a_3 a_4 ]}^{c_2 c_3 ]d} +\delta_b^d P_{[a_1 }\delta_{a_2a_3a_4 ]}^{c_1c_2 c_3})  \; .
\eqno(2.4.41)$$
\medskip
{\bf 3 The Cartan forms}
\medskip
\noindent
As explained in section one the non-linear realisation, and so the dynamics,  is constructed from the Cartan forms. Given the $E_{11}\otimes_s l_1$ algebra constructed previously in this paper these are easily found. We write the group element of $E_{11}\otimes_s l_1$ in the form   $g = g_lg_E$ where

$$ g_E  = \ldots e^{A_{a_1\ldots a_5, b PQ}R^{a_1\ldots a_5,b PQ}}e^{A_{a_1\ldots a_6 (PQ)}R^{a_1\ldots a_6(PQ)}}e^{A_{a_1\ldots a_6}{}^{PQ,R}R^{a_1\ldots a_6}{}_{PQ,R}} $$ 
$$ e^{h_{a_1\ldots a_4,b}R^{a_1\ldots a_4,b}}e^{\varphi_{a_1\ldots a_5 M}{}^NR^{a_1\ldots a_5 M}{}_N}e^{A_{a_1\ldots a_4}{}^{MN}R^{a_1\ldots a_4}{}_{MN}} $$ $$ e^{A_{a_1a_2a_3M}R^{a_1a_2a_3M}}e^{A_{a_1a_2}{}^MR^{a_1a_2}{}_M}e^{A_{aMN}R^{aMN}}e^{\varphi_M{}^NR^M{}_N}e^{h_a{}^bK^a{}_b} \; , \eqno(3.1) $$

\noindent
and

$$ g_l = e^{x^aP_a}e^{x_{MN}Z^{MN}}e^{x_a{}^MZ^a{}_M}e^{x_{a_1a_2M}Z^{a_1a_2M}}e^{x_{a_1a_2a_3}{}^{MN}Z^{a_1a_2a_3}{}_{MN}} $$
$$ e^{x_{a_1a_2a_3,b}Z^{a_1a_2a_3,b}}e^{x_{a_1\ldots a_5}Z^{a_1\ldots a_4}}e^{x_{a_1\ldots a_4M}{}^NZ^{a_1\ldots a_5M}{}_N} \; . \eqno(3.2) $$
\par
We note that the parameters of the group element  $g_E$ will be the fields of the theory while those of the group element $g_l$ correspond to the generalised spacetime coordinates. The fields depend on the coordinates. In the above form of the group element we have gauged away the part of the $E_{11}$ group element that depends on the negative level generators using the local $I_c(E_{11})$ symmetry. This symmetry also contains the local Lorentz symmetry and we have not used this to make the graviton field symmetric. The analogous statement holds for the internal SO(1,4) symmetry.
\par
The Cartan form of  equation (1.5) can be written as
$$ {\cal V} = G_a{}^bK^a{}_b + G_M{}^NR^M{}_N + G_{aMN}R^{aMN} + G_{a_1a_2}{}^MR^{a_1a_2}{}_M $$
$$ + G_{a_1a_2a_3M}R^{a_1a_2a_3M} + G_{a_1\ldots a_4}{}^{MN}R^{a_1\ldots a_4}{}_{MN} $$
$$ + G_{a_1\ldots a_4,b}R^{a_1\ldots a_4,b}+G_{a_1\ldots a_5 M}{}^NR^{a_1\ldots a_5 M}{}_N  + G_{a_1\ldots a_5, b PQ}R^{a_1\ldots a_5,b PQ} $$ $$ + G_{a_1\ldots a_6 (PQ)}R^{a_1\ldots a_6(PQ)} + G_{a_1\ldots a_6}{}^{PQ,R}R^{a_1\ldots a_6}{}_{PQ,R} + \ldots \; .  \eqno(3.3)$$
In this paper, we only need the  Cartan forms up to  level 6.
\par
The explicit for of the G's in terms of the fields of equation (3.1) up to level 4 are
$$ G_a{}^b = (e^{-1}de)_a{}^b \; , $$
$$ G_M{}^N = (f^{-1}df)_M{}^N \; , $$
$$ G_{aMN} = e_a{}^{\mu}f_{M}{}^{\dot M}f_N{}^{\dot N}d A_{\mu \dot M \dot N} \; , $$
$$ G_{a_1a_2}{}^M = e_{a_1}{}^{\mu_1}e_{a_2}{}^{\mu_2}f^{ M}{}_{\dot M}(dA_{\mu_1\mu_2}{}^{\dot M} - {1\over 2} \varepsilon^{\dot M \dot N \dot P \dot Q \dot R} A_{[\mu_1 \dot N \dot P}dA_{\mu_2] \dot Q \dot R})  \; , $$
$$ G_{a_1a_2a_3M} = e_{a_1}{}^{\mu_1}e_{a_2}{}^{\mu_2}e_{a_3}{}^{\mu_3}f_M{}^{\dot M}(dA_{\mu_1\mu_2\mu_3 \dot M} - A_{[\mu_1 \dot N \dot M} dA_{\mu_2\mu_3]}{}^{\dot N} $$
$$ + {1\over 3!}A_{[\mu_1 \dot N \dot M}A_{\mu_2 \dot R \dot S} dA_{\mu_3]\dot P \dot Q} \varepsilon^{\dot P \dot Q \dot R \dot S \dot N}) \; , $$
$$ G_{a_1\ldots a_4}{}^{MN} = e_{a_1}{}^{\mu_1}\ldots e_{a_4}{}^{\mu_4}f^{M}{}_{\dot M}f^N{}_{\dot N}( dA_{\mu_1\ldots \mu_4}{}^{\dot M \dot N} $$
$$ - \varepsilon^{\dot M \dot N \dot P \dot Q \dot R}A_{[\mu_1 \dot P \dot Q} dA_{\mu_2\mu_3\mu_4]\dot R} - {1\over 2}\varepsilon^{\dot M \dot N \dot P \dot Q \dot R} A_{[\mu_1\dot P \dot Q} A_{\mu_2 \dot R \dot S}dA_{\mu_3\mu_4]}{}^{\dot S}  $$
$$ - 5 A_{[\mu_1 \dot P \dot Q}A_{\mu_2}{}^{[\dot R \dot Q}A_{\mu_3 }{}^{\dot P}{}_{\dot R}dA_{\mu_4]}{}^{\dot M \dot N]}) \; . \eqno(3.4) $$
where $e_{\mu}{}^a = (e^{h})_{\mu}{}^a$ and $ f_{\dot M}{}^{ N} = (e^{\varphi})_{\dot M}{}^{ N}   $.
The dotted latin indices $\dot M, \dot N, \ldots $ can be thought of as curved indices and the $M,N,\ldots $ indices as flat indices in the internal space. Indeed we can regard $f^{N}{}_{\dot M}$ as the analogue of usual vielbein $e_\mu{}^a$ but in the internal space which has the local SO(1,4) symmetry.
\par
We observe that for  the gauge choice in the local $I_c(E_{11})$ symmetry  in which we are working the Cartan forms corresponding to negative level generators vanish,  as was implicitly taken account of when we wrote equation (3.3).
\medskip
{\bf 4 Transformations of the Cartan forms }
\medskip
To find the equations of motion we will need the transformations of the  Cartan forms under the local transformations as in equation (1.4).
We recall that these local transformations act on the Cartan forms as in equation (1.8) and that the Cartan forms  are inert under the rigid
$E_{11}\otimes_s l_1$  transformations. The local transformation $h \in I_c(E_{11})$ which involves the $\pm 1$ generators of $E_{11}$  is of the form
$$ h = 1 - \Lambda_{aMN} S^{aMN} ,  \eqno(4.1) $$
where $ S^{aMN} = R^{aMN} - R_{aMN} $. If the equations of motion are invariant under these transformations and the local $SO(1,4)\times SO(5)$ transformations at level zero then they are invariant under the full $I_c(E_{11})$ transformations.
\par
We find that under the above local transformation, the Cartan form transforms as

$$ \delta {\cal V}_E = [\Lambda_{aMN}S^{aMN}, {\cal V}_E] - S^{aMN}d\Lambda_{aMN} ,  \eqno(4.2)$$
Evaluating this equation using the algebra given in section three we  find that   the Cartan form transformations up to level 5,  as
$$ \delta G_a{}^b = 2 \Lambda^{aMN}G_{bMN} - {2\over 5} \delta^a_b \Lambda^{cMN}G_{cMN} \; ,
\eqno(4.3) $$
$$ \delta G_M{}^N = 4 \Lambda^{cPN}G_{cPM} - {4\over 5}\delta_M^N \Lambda^{cPQ}G_{cPQ}
\eqno(4.4) $$
$$ \delta G_{aMN} = - \Lambda_{bMN}G_{a}{}^{b} - 2\Lambda_{aP[N}G_{M]}{}^P - \varepsilon_{MNPQR}\Lambda_{bQR}G_{ba}{}^P - d\Lambda_{aMN} \; ,
\eqno(4.5) $$
$$ \delta G_{a_1a_2}{}^M = \varepsilon^{MNPQR}\Lambda_{[a_1NP}G_{a_2]QR} + 12 \Lambda^{bNM} G_{ba_1a_2N}                         \; ,
\eqno(4.6) $$
$$ \delta G_{a_1a_2a_3M} = \Lambda_{[a_1NM}G_{a_2a_3]}{}^N - 2 \varepsilon_{MPQRS}\Lambda_{bPQ}G_{ba_1a_2a_3}{}^{RS}                             \; ,
\eqno(4.7) $$
$$ \delta G_{a_1\ldots a_4}{}^{MN} = \varepsilon^{MNPQR}\Lambda_{[a_1PQ}G_{a_2a_3a_4]R} + 20 \Lambda^{bP[N}G_{ba_1\ldots a_4P]}{}^{M]} $$
$$ - 2 \Lambda_{bMN}G_{a_1\ldots a_4,b}  \; ,
\eqno(4.8) $$
$$ \delta G_{a_1\ldots a_5 M}{}^N = - 2 \Lambda_{[a_1PM}G_{a_2\ldots a_5]}{}^{PN}+ {2\over 5} \delta_M^N \Lambda_{[a_1PQ}G_{a_2\ldots a_5]}{}^{PQ} $$ $$ + {2\over 5} \Lambda^{dNP} G_{a_1\ldots a_5,dMP} - {2 \over 25} \delta^N_M \Lambda^{dPQ}G_{a_1\ldots a_5,d PQ} $$ $$ + 2 \Lambda^{dPQ}\varepsilon_{PQRSM}G_{da_1\ldots a_5}{}^{S(N,R)} + 3 \Lambda^{dNP}G_{da_1\ldots a_5(MP)}
 \; ,
\eqno(4.9) $$
$$ \delta G_{a_1\ldots a_4,b} = {4\over 5} \Lambda_{bMN}G_{a_1\ldots a_4}{}^{MN} - {4\over 5} \Lambda_{[a_1MN}G_{a_2\ldots a_4]b}{}^{MN} $$ $$ - \Lambda^{cMN}G_{ca_1\ldots a_4,b MN} + \Lambda^{cMN}G_{c[a_1\ldots a_4,b]MN}\; ,
 \eqno(4.10) $$
 $$ \delta G_{a_1\ldots a_6}{}^{MN,P} = 2 \varepsilon^{QRSMN}\Lambda_{[a_1 QR} G_{a_2\ldots a_6]S}{}^P - 2 \varepsilon^{QRS[MN}\Lambda_{[a_1QR} G_{a_2\ldots a_6]T}{}^{P]} + \ldots \; ,
 \eqno(4.11) $$
 $$ \delta G_{a_1\ldots a_6(MN)} = 8 \Lambda_{[a_1 (N|P} G_{a_2\ldots a_6]|M)}{}^P + \ldots \; ,
 \eqno(4.12) $$
 $$ \delta G_{a_1\ldots a_5, b MN} = 20 \Lambda_{b[N|P}G_{a_1\ldots a_5 |M]}{}^P $$
 $$ - 20 \Lambda_{[b[N|P}G_{a_1\ldots a_5]|M]}{}^P + 10 \Lambda_{[a_1MN}G_{a_2\ldots a_5],b} + \ldots \; .
 \eqno(4.13) $$
where the $\ldots$ represent level 7 forms.
\par
In fact the first term of equation (4.2) does not preserve our gauge choice as it contains terms which contain the level minus one generators. To correct for this we must compensate by choosing the second term in equation (4.2) so that we do preserve our gauge choice. In particular following the treatment of the eleven dimensional theory [7,8],  we must require the condition
$$ [\Lambda \cdot R^{(-1)}, {\cal V}^{(0)}] - d\Lambda \cdot R^{(-1)} = 0 \; ,
 \eqno(4.14) $$
where the superscripts give the relevant level.
This equation implies that  the $\Lambda^{aMN}$ must obey the equation
$$ d\Lambda^{aMN} = G_{b}{}^a\Lambda^{bMN} + 2 G_P{}^{[M}\Lambda^{a|P|N] } \; , \eqno(4.15) $$
which  is solved by
$$ \Lambda_{c}{}^{\mu \dot K \dot L} e_{\mu}{}^af_{\dot K}{}^Mf_{\dot L}{}^N = \Lambda^{aMN} \; , \eqno(4.16) $$
where the $c$ subscript represents the  fact that the $\Lambda^{\mu \dot K \dot L}$ with curved indices is a constant. Substituting  this  condition back into the transformation of the level 1 Cartan form in equation (4.5)  to find that
$$ \delta G_{aMN} = - 2\Lambda_{bMN}G_{(ba)} - 4\Lambda_{aP[N}G_{(M]P)} - \varepsilon_{MNPQR}\Lambda_{bQR}G_{ba}{}^P \eqno(4.17)$$

\noindent
We  note at this point, that in the following calculations, we come across the derivative of this parameter $\Lambda^{aMN}$, and that using equation (4.16), we notice that only $\Lambda^{\mu \dot K \dot L}$ with curved indices is independent of the generalised space-time coordinates.
\par
So far we have written the Cartan forms as forms and so what we actually have written say in the last section is
$$ G_{\underline \alpha} = dz^{\Pi}G_{\Pi, \underline \alpha} \; , \eqno(4.18) $$
where the  index  $\Pi$ represents the $l_1$ representation, and  index $\underline \alpha$ is an $E_{11}$ index. We notice that once the Cartan forms are written in this way, they are no longer invariant under the rigid transformations. We can correct this by changing the first index into a tangent index using the inverse vielbein
$$ G_{A,\underline \alpha} = (E^{-1})_A{}^{\Pi}G_{\Pi, \underline \alpha} \; , \eqno(4.19) $$
and then the  Cartan forms with the local flat index are inert under the rigid transformations.  We will refer to this first index as the $l_1$ index.
\par
We find, using equation (1.9),  that the variation on the $l_1$ tangent index under the above local transformation is given by
$$\delta G_{a, \alpha} = - 2\Lambda_{aMN}G^{MN}{}_{,\alpha} \; , \eqno(4.20)$$

$$ \delta G^{MN}{}_{, \alpha} = \Lambda^{aMN}G_{a,\alpha} \; . \eqno(4.21)$$
Of course the $E_{11}$ index also transforms as in equations (4.3-4.13) and (4.17) and so the actual local $I_c(E_{11})$ transformation of the above Cartan form is the sum of the terms in the two equations.
\par
We conclude this section by making some observations that will be essential when calculating the equations of motion in the next section.
Let us consider the transformation of the level one Cartan form $G_{a_1,a_2 MN}$ of equation (4.17). Even we antisymmetrise  the $a_1$ and $a_2$ indices we observe that the transformation contains Cartan forms that do not have their indices antisymmetrised in a similar way, that is, the indices  involving the $l_1$ index. We will find that the calculation of the equations of motion requires objects for which  the indices are antisymmetrised. We will now show how to construct objects with this property by using the coordinates beyond those of spacetime which are in the vector representation. Such a level one  object is given by
$$ {\cal G}_{a_1a_2 MN} = G_{[a_1,a_2] MN} + \varepsilon_{MNPQR} G^{QR}{}_{,a_1a_2}{}^P \; , \eqno(4.22) $$
which  transforms as
$$ \delta {\cal G}_{a_1 a_2 MN} = - 2\Lambda^{bMN}G_{[a_1,(a_2]b)} - 4\Lambda_{[a_2 P[N}G_{a_1],(M]P)} - {3\over 2} \varepsilon_{MNPQR}\Lambda_{bQR} G_{[a_1,ba_2]}{}^P \; ,  \eqno(4.23) $$
where we have kept only terms that involve derivatives with respect to the usual coordinates of spacetime. We observe that the right-hand side contains Cartan forms whose $l_1$ index is antisymmetrised with its  spacetime $E_{11}$ indices.
\par
The corresponding object at level two is given by
$$ {\cal G}_{a_1a_2a_3}{}^M = G_{[a_1,a_2a_3]}{}^M - 4G^{NM}{}_{,a_1a_2a_3N} \; , \eqno(4.24) $$
It has the local transformation
$$ \delta {\cal G}_{a_1a_2a_3}{}^M = \varepsilon^{PQRSM}\Lambda_{[a_2 PQ} G_{a_1,a_3]RS} + 16 \Lambda_{bNM} G_{[a_1,ba_2a_3]N} \; . \eqno(4.25) $$
\par
At level 3 we define the object
$$ {\cal G}_{a_1\ldots a_4 M} = G_{[a_1,a_2a_3a_4] M} + {1 \over 2}\varepsilon_{PQRSM}G^{PQ}{}_{,a_1a_2a_3a_4 RS} \; , \eqno(4.26)$$
It  has the transformation
$$ \delta {\cal G}_{a_1\ldots a_4 M} = \Lambda_{[a_2NM}G_{a_1,a_3a_4]}{}^N - {5\over 2} \Lambda_{bPQ}\varepsilon_{PQRSM}G_{[a_1,ba_2a_3a_4] RS} \; . \eqno(4.27) $$
\par
At level 4, we define
$$ {\cal G}_{a_1\ldots a_5}{}^{MN} = G_{[a_1,\ldots a_5]}{}^{MN} + 4 G^{P[N}{}_{,a_1\ldots a_5 P}{}^{M]} \; , \eqno(4.28) $$
which has the transformation
$$ \delta {\cal G}_{a_1\ldots a_5}{}^{MN} = \varepsilon_{PQRMN}\Lambda_{[a_2 PQ}G_{a_1,a_3a_4a_5]R} $$
$$ + 24 \Lambda^{bP[N}G_{[a_1,ba_2\ldots a_5] P}{}^{M]}  - 2 \Lambda_{bMN}G_{[a_1, \ldots a_5],b} \; . \eqno(4.29)  $$

At level 5, we define 
$$ {\cal G}_{[a_1,\ldots a_6]M}{}^{N} = G_{[a_1,\ldots a_6]M}{}^{N} - {1\over 3} \varepsilon_{PQRSM}G^{PQ}{}_{,a_1\ldots a_6}{}^{S(N,R)} - {1\over2}G^{NP}{}_{, a_1\ldots a_6 (MP)} \; ,
\eqno(4.30)$$
and
$$ {\cal G}_{[a_1,\ldots a_5],b} = G_{[a_1,\ldots a5],b} + {3 \over 25} G^{MN}{}_{,a_1\ldots a_5,bMN} \;,
\eqno(4.31)$$
which transform as
$$ \delta G_{[a_1, \ldots a_6] M}{}^{N} = - 2 \Lambda_{[a_2PM} G_{a_1,a_3 \ldots a_6]}{}^{PN} + {2\over 5}\delta^N_M \Lambda_{[a_2 PQ}G_{a_1,a_3 \ldots a_6]}{}^{PQ} $$ $$- {2\over5} \Lambda^{dNP}G_{[a_1,\ldots a_6],d MN} - {2\over 25} \delta^N_M \Lambda^{dPQ} G_{[a_1,\ldots a_6],dPQ} $$ $$+ {7\over 3} \Lambda^{dPQ} \varepsilon_{PQRSM} G_{[a_1,da_2\ldots a_6]}{}^{S(N,R)} + {7 \over 2} \Lambda^{dNP} G_{[a_1,da_2\ldots a_6] (MP)}  \; , 
\eqno(4.32) $$
and
$$ \delta G_{[a_1,\ldots a_5],b} = {4\over5} \Lambda_{bMN}G_{[a_1,\ldots a_5]}{}^{MN} - {4\over 5} \Lambda_{[a_2MN} G_{a_1,a_3a_4a_5]b}{}^{MN} $$ $$ - {6\over 5 } \Lambda^{cMN}G_{a_1,\ldots a_5c],bMN} + {6 \over 25} G_{[a_1,\ldots a_5,b],cMN}  \; , 
\eqno(4.33) $$
\par
Finally, at level 6, we find the transformations
$$ \delta G_{[a_1,\ldots a_7]}{}^{MN,P} = 2 \varepsilon^{QRSMN} \Lambda_{[a_2 QR}G_{a_1,a_3\ldots a_7]S}{}^P $$ $$ -  2 \varepsilon^{QRS[MN} \Lambda_{[a_2 QR}G_{a_1,a_3\ldots a_7]S}{}^{P]} + \ldots \; ,
\eqno(4.34) $$
$$ \delta G_{[a_1,\ldots a_7] (MN)} = 8 \Lambda_{[a_2(N|P}G_{a_1,a_3\ldots a_7]|M)}{}^{P} + \ldots \; ,
\eqno(4.35) $$
$$ \delta G_{[a_1,\ldots a_6],b MN} = {50 \over 3} \Lambda_{b[N|P} G_{[a_1,\ldots a_6]|M]}{}^P $$ $$ + {50 \over 3} \Lambda_{[a_2[N|P} G_{a_1,a_3\ldots a_6]b|M]}{}^P + 16 \Lambda_{[a_2MN} G_{a_1,a_3\ldots a_6],b} + \ldots \; ,
\eqno(4.36) $$
where the $\ldots$ represent higher level Cartan forms. We note that we have not defined new $l_1$ modified objects at level 6, as we are only concerned with the transformation into the lower level Cartan forms in this paper and these additions would involve level seven fields.
\par
One can think about the above definition in terms of gauge symmetry which requires that the indices be antisymmetric. As a result one see that the extra coordinates beyond those of spacetime are required to ensure gauge symmetry. In the $E_{11}$ approach we do not require gauge symmetries only the symmetries of the non-linear realisation. It turns out that the results are indeed gauge invariant!
\medskip
{\bf 5 The  equations of motion }
\medskip
We begin the calculation by finding  duality relations for the form fields contained in the theory. We can expect that the one form is dual to the four form and if this were the case it would obey an equation of the form
$$ D{}_{a_1a_2 MN}\equiv {\cal G}_{a_1a_2MN} + e_2 \varepsilon_{a_1a_2}{}^{b_1\ldots b_5}{\cal G}_{b_1\ldots b_5}{}^{MN}=0 \; , \eqno(5.1) $$
Simlarly we may expect that the two form is dual to the three form and so this equation would be of the form
$$ D{}_{a_1a_2a_3}{}^{ M} \equiv  {\cal G}_{a_1a_2a_3}{}^M + e_3 \varepsilon_{a_1a_2a_3}{}^{b_1\ldots b_4} {\cal G}_{b_1\ldots b_4 M }=0 \; . \eqno(5.2) $$
where $e_2$ and $e_3$ are constants. The equation are restricted to be of the form on the grounds of Lorentz symmetry. To see if the theory really does have these duality equation we must examine if they are invariant under the local $I_c(E_{11})$ transformations of
equations (4.3)-(4.13) and equation (4.17). One finds that the equations vary into each other under these transformations if
 the constants $e_2$ and $e_3$ take the values  $e_2 = \mp {i \over 2}$ and  $e_3 = \pm {i \over 3} $. In the following, we choose to work with the first sign, i.e.  $e_2 = -{i\over 2}$ and $ e_3 = {i\over 3}$, but one can easily recover where the sign changes for the other case (simply by changing the sign whenever a $i$ appears).  The precise transformations are
$$
\delta {{ D}}{}_{a_1a_2 MN} = -{3\over 2} \varepsilon_{MNPQR}\Lambda_{bQR}{D}_{a_1ba_2}{}^P + \dots  \; , \eqno(5.3) $$
$$
\delta {D}{}_{a_1a_2a_3 M} = \varepsilon^{PQRSM}\Lambda_{[a_2PQ}{{ D}}_{a_1a_3]RS} + {i \over 3} \varepsilon_{a_1a_2a_3}{}^{b_1\ldots b_4} \Lambda_{b_2 NM}{D}_{b_1b_3b_4}{}^N \; . \eqno(5.4) $$
where $+\ldots$ means the addition of  terms involving the gravity  and the scalars  fields.
As anticipated in section four the invariance of the equations requires the objects defined in that section that have totally antisymmetrised indices. Roughly speaking the equations contain objects with such indices and so their variations must also contain  objects  with such indices in order for a cancelation to take place. We could have started out with the usual Cartan forms but then we would have found that we needed to add terms which was equivalent to using the above objects. The fact that the duality relations are invariant up to scalar and gravity terms tells us that these are the correct equations. The scalar and gravity terms will be cancelled by terms in the variations that are the gravity and scalar equations of motion. We will return to the full variation of the duality conditions in a future paper.
\par
We now proceed as was done in the  [78], namely we will work with equations that are second order in derivatives and only contain the gravity field and the one and two form fields. To this end we take the derivatives of   the equations (5.1) and (5.2) in such a way that
the three form and four form field drop out respectively.   In doing this we need to use the explicit forms of the Cartan forms given in section three. Carrying this out on  ${\cal D}$ of   equation (5.1) we find the result

$$ \partial_{\mu_1}(\det(e)^{1\over 2} G^{[\mu_1,\mu_2] \dot M \dot N}) + {2\over 3} G_{[\mu_1, \mu_3]\dot Q \dot R}G^{[\mu_1,\mu_2\mu_3]}{}_{ \dot P} \varepsilon^{\dot M \dot N \dot P \dot Q \dot R}  $$
$$ + {i \over 4} \varepsilon^{\mu_1\mu_2 \nu_1 \ldots \nu_5} G_{[\mu_1, \nu_2\nu_3]}{}^{[\dot M}G_{[\nu_1,\nu_4\nu_5]}{}^{\dot N]} = 0 \; , \eqno(5.5)$$
\par
The equivalent  operation on the duality relation of  equation (5.2) leads to the equation
$$ \partial_{\mu_1}(\det(e)^{1\over 2} G^{[\mu_1,\mu_2\mu_3] }{}_{\dot P}) - {i \over 3}\varepsilon^{\mu_1\mu_2\mu_3 \nu_1 \ldots \nu_4}G_{[\mu_1, \nu_2] \dot N \dot P}G_{[\nu_1,\nu_3\nu_4]}{}^{\dot N} = 0  \; .
 \eqno(5.6) $$
 We notice that the factors of the $\det(e) ^{1\over 2}$ appear due to their appearance in the inverse vielbein, a feature that is also present in the  equivalent eleven dimensional calculation [7,8].
\par
In order to vary equations (5.5) and (5.6) we will need to rewrite them in terms of the Cartan forms of section three whose variation we found in section four. Essentially this means rewriting it in such a way that all the indices on the Cartan forms are tangent indices both for spacetime as well as internal indices.  This proceedure leads to the two equations
$$ E{}^ {a_2 QR} = {1\over 2} G_{a_1,d}{}^d G^{[a_1,a_2]}{}^{QR} - G_{a_1,d}{}^{a_1}G^{[d,a_2]}{}^{QR} $$
$$ - G_{a_1,d}{}^{a_2}G^{[a_1,d]}{}^{QR} - 2 G_{a_1,P}{}^{[Q|}G^{[a_1,a_2]}{}^{P|R]} + \det(e)^{1\over 2} e_{a_1}{}^{\mu}\partial_{\mu}(G^{[a_1,a_2]}{}^{QR}) $$
$$ + {2\over 3} G_{[a_1, a_3] MN }G^{[a_1,a_2a_3]}{}_{  P} \varepsilon^{ M  N  P  Q  R}  $$
$$ + {i \over 4} \varepsilon^{a_1a_2 b_1 \ldots b_5} G_{[a_1, b_2b_3]}{}^{[ Q}G_{[b_1,b_4b_5]}{}^{R]} = 0 \; , \eqno(5.7)$$
and
$$ E{}^{ a_2a_3}{}_{ M} = {1\over 2} G_{a_1,d}{}^d G^{[a_1,a_2a_3]}{}_{M} - G_{a_1,d}{}^{a_1}G^{[d,a_2a_3]}{}_{M} $$
$$ - 2 G_{a_1,d}{}^{[a_2|}G^{[a_1,d|a_3]]}{}_{M} + G_{a_1,M}{}^P G^{[a_1,a_2a_3]}{}_P + \det(e)^{1\over 2} e_{a_1}{}^{\mu} \partial_{\mu}(G^{[a_1,a_2a_3]}{}_{M}) $$
$$ -  {i \over 3}\varepsilon^{a_1a_2a_3 b_1 \ldots b_4}G_{[a_1, a_2]  N  P}G_{[b_1,b_3b_4]}{}^{ N} = 0 \; , \eqno(5.8) $$
Although we have derived the above equations from the duality relations that are first order in derivatives we can regard these last two  equations as the starting point of our quest to find the equations of motion for the fields in terms of which the seven dimensional theory is usually written. We will now vary them under the $I_c(E_{11})$ transformations and show that, together  with some other equations of motion, they are invariant.  In carrying out this calculation we have neglected all terms that contain derivatives with respect to the higher level coordinates for reasons we will explain shortly.
\par
Let us suppose that in the variation of one of the equations of motion we have a term of the form
$$ \Lambda^{dMN}G_{d, \alpha}f^{\alpha}{}_{MN} \; ,
\eqno(5.9) $$
where $f^{\alpha}{}_{MN}$ is any function of the fields and their derivatives. Then we can cancel such a term by adding to the equations of motion the term
$$ - G^{MN}{}_{, \alpha}f^{\alpha}{}_{MN} \; ,
\eqno(5.10) $$
We will call such terms $l_1$ terms. We carry out our variations keeping only terms that contain derivatives with respect to the usual coordinates of spacetime. However, the last remark implies that to do this we must find the equations of motion up to the level that contains all terms that have the usual spacetime derivatives as well as terms that  that are first order in derivatives with respect to the level one coordinates, that is, the $l_1$ terms. Even if we begin with an equation of motion, such as those in equations (5.7) and (5.8), we will find the $l_1$ terms. However, we do not find the terms that contain derivatives with respect coordinates of level two and above.
We note that in finding the duality relations of equations (5.1) and (5.2) we did find the $l_1$ terms. In this way of doing things
we do not for example keep the variation of the term in equation (5.10) that contains derivative with respect to the level two coordinates.
\par
We will now carry out the local $I_c(E_{11})$ transformations of equation (5.7).
If we vary the terms in the first two lines, we the result
$$ e_{a_2}{}^{\mu_2} \partial_{\mu_1}( \omega_{\tau, \mu_1\mu_2} \det(e) - \det (e)^{1 \over 2} G_{\tau, [\mu_1\mu_2]}) \Lambda_{\tau QR} $$
$$ - 2 \Lambda^{a_2 MN} G_{a_1, d MN} G_{[a_1,d]}{}^{QR} - 8 \Lambda^{c P[Q|}G_{a_1,c PM} G_{[a_1,a_2]}{}^{M |R]} $$
$$ - 2\Lambda^{a_1 MN} G_{a_1,dMN} G_{[d,a_2]}{}^{QR} + 2 \Lambda ^{dMN}G_{a_1,d MN} G_{[a_1,a_2]}{}^{QR} \; , \eqno(5.11)$$
where
$$ \det(e){}^{1 \over 2} w_{c,ab} = - G_{a,(bc)} + G_{b,(ac)} + G_{c,[ab]} \; . \eqno(5.12) $$
\par
We note that the terms on the second line and also the last term will be cancelled with terms from the transformation of the final terms in equation (5.7). We also notice that the second  and the fifth  terms are terms of the form given in equation (5.9), and so we can cancel them by adding corresponding $l_1$ terms. Finally, we can manipulate the first term in the following way. We notice that
$$ e_{\mu}{}^a \partial_{\nu} ( \det(e) \omega_{\tau,}{}^{\nu\mu}) = $$
$$ \det(e)(e_b{}^{\nu}\partial_{\nu} \omega_{\tau,}{}^{ba} + (e_{\mu}{}^a\partial_{\nu}e_c{}^{\mu})\omega_{\tau,}{}^{\nu c} + (e_c{}^{\lambda}\partial_{\nu}e_{\lambda}{}^{c}) \omega_{\tau,}{}^{\nu a} + \partial_{\nu} e_b{}^{\nu}\omega_{\tau, }{}^{ba}) \eqno(5.13) $$
The first term will turn out to just what we will need.  While the second term we can be rewriten as
$$ e_b{}^{\lambda}\partial_{\nu}e_{\lambda}{}^a\omega_{\mu ,}{}^{\nu b} = G_{c,\nu a}\omega_{\mu , }{}^{c \nu} = (-G_{a,(c\nu )} + G_{c,(a\nu )} + G_{\nu , [ac]} ) \omega_{\mu ,}{}^{c\nu} = \omega_{\nu ,}{}^a{}_c \omega_{\mu ,}{}^{c\nu} \; , \eqno(5.14) $$
and the final two terms in (5.13) can be written as
$$ \omega_{\mu,}{}^{ab}\partial_{\lambda}(\det(e)e_b{}^{\lambda}) = - \det(e) \omega_{\mu,}{}^{ab}\omega_{\lambda ,}{}^{b \lambda} \; . \eqno(5.15) $$
We note that the Ricci tensor is given by
$$ R_{\mu}{}^a = \partial_{\mu}\omega_{\nu,}{}^{ab} e_b{}^{\nu} - \partial_{\nu}\omega_{\mu,}{}^{ab}e_b{}^{\nu} + \omega_{\mu,}{}^{a}{}_c\omega{}_{\nu,}{}^{cb}e_b{}^{\nu} - \omega_{\nu,}{}^a_c \omega_{\mu,}{}^{cb}e_b{}^{\nu} \; , \eqno(5.16) $$
and so in  equation (5.13) we find the expression
$$ \det(e)(R_{\tau}{}^{a_2} - \partial_{\tau}(\omega_{\nu,} {}^{a_2b}) e_{b}{}^{\nu}) \Lambda^{\tau QR} \; , \eqno(5.17) $$
and again the second  term in this is an $l_1$ term of the form of equation (5.9).
\par
Including $l_1$ terms, we find that the equation of motion for the one form is
 $$ {\cal E}{}^{ a_2 QR} \equiv {1\over 2} G_{a_1,d}{}^{d} G_{[a_1,a_2]}{}^{QR} - G_{a_1,d}{}^{a_1}G_{[d,a_2]}{}^{QR} $$
$$ - G_{a_1,d}{}^{a_2}G_{[a_1,d]}{}^{QR} - 2 G_{a_1,P}{}^{[Q|}G_{[a_1,a_2]}{}^{P|R]} + \det(e)^{1\over 2} e_{a_1}{}^{\mu}\partial_{\mu}(G_{[a_1,a_2]}{}^{QR}) $$
$$ +{3\over 2}G_{[a_1,a_3]MN} G_{[a_1,a_2a_3]P} \varepsilon^{MNPQR} + {i \over 4} \varepsilon^{a_1a_2b_1 \ldots b_5}  G_{[a_1,b_2b_3]}{}^{[Q}G_{[b_1,b_4b_5]}{}^{R]}  $$
$$ + 8 G^{P[Q}{}_{,a_1 PM}G_{[a_1,a_2]}{}^{M|R]} + e_{a_2}{}^{\mu_2}\partial_{\mu_1}(\det(e)^{1\over 2}G^{QR}{}_{,[\mu_1\mu_2]}) $$
$$ + \det(e)\partial^{QR}(\omega_{\nu},{}^{a_2b})e_b{}^{\nu} - (\det(e)^{1\over2}\partial^{P[R|}(G_{a_2}{}_{,|Q]}{}^{P} +  G_{a_2}{}_{,P}{}^{|Q]})  $$
$$ + {1\over2}(G_{a_2}{}_{,|Q]}{}^{P}+G_{a_2}{}_{,P}{}^{|Q]})G^{P[R|}{}_{,d}{}^d $$
$$ -( G_{a_2}{}_{,N}{}^P + G_{a_2}{}_{,P}{}^N )G^{P[R|}{}_{,N}{}^{|Q]}  + ( G_{a_2}{}_{,|Q]}{}^{N} +  G_{a_2}{}_{,N}{}^{|Q]}) G^{P[R|}{}_{,P}{}^N $$
$$ + (G_{d}{}_{,|Q]}{}^{P} + G_{d}{}_{,P}{}^{|Q]})G^{P[R|}{}_{,d}{}^{a_2}) =0\; .  \eqno(5.18)$$
\par
Using  the same reasoning, we also find that the equation of motion for the two form is
$$ {\cal E}{}^{ a_2a_3 M} \equiv {1\over 2} G_{a_1,d}{}^{d} G^{[a_1,a_2a_3]M} - G_{a_1,d}{}^{a_1}G^{[d,a_2a_3]M} $$
$$ - 2 G_{a_1,d}{}^{[a_2|}G^{[a_1,d|a_3]]M} + G_{a_1,M}{}^P G_{[a_1,\ldots a_3]P} + \det (e)^{1 \over 2} e_{a_1}{}^{\mu}\partial_{\mu}(G^{[a_1,a_2a_3] M}) $$
$$ - {i \over 3} \varepsilon ^{a_1\ldots a_3 b_1\ldots b_4}(G_{[a_1,b_2]NM} G_{[b_1,b_3b_4]}{}^{N}) $$
$$ +{1\over 3}\varepsilon^{PQRSM}({1\over 2} G^{PQ}{}_{,d}{}^d G_{[a_2,a_3]RS} - 2 G^{PQ}{}_{,d}{}^{[a_2|}G^{[d,|a_3]] RS} + \det(e)^{1\over 2}\partial^{PQ}(G_{[a_2,a_3]RS})) $$
 $$ - {2\over 3} G^{QR}{}_{,S}{}^{[P|}G_{[a_2,a_3]}{}^{S|T]}\varepsilon^{TMPQR} + 2 G^{PQ}{}_{,dPQ}G_{[d,a_2a_3]M} - 4 G^{NP}{}_{,a_1NM}G_{[a_1,a_2a_3]P}=0  \; .  \eqno(5.19) $$
\par
The  transformation of the one form equation of motion is given by
$$ \delta {\cal E}{}^{ a_2 QR} =  -{3\over 2}\Lambda _{cMN}\varepsilon^{MNPQR}{\cal E}{}^{ca_2}{}_P + \Lambda^{bQR}E_{b}{}^{a_2} + \Lambda_{a_2 P[Q|}{\cal E}_{(|R]P)} $$
$$ - {i \over 2} \varepsilon^{a_1a_2b_1\ldots b_5}\varepsilon^{MNPQR} G_{[a_1,b_2]MN}\Lambda_{[b_3TP}{ D}_{b_1,b_4b_5]}{}^T  + 4! \Lambda_{[a_4}{}^{P [R}{\cal D}{}_{a_1a_2a_3]P}G_{[a_1,a_3a_4]}{}^{ Q]} \; , \eqno(5.20) $$
where $ E_{a}{}^b $ is
$$ E_{a}{}^b = \det(e) R_a{}^b - G_{a,(MN)}G_{b,(MN)} $$
$$ - 2 (2 G_{[c,a]}{}^{MN}G_{[c,b]MN} - {1\over 5} \delta^b_a G_{[c_1,c_2]}{}^{MN}G_{[c_1,c_2]MN}) $$
$$ - 3(3 G_{[c_1,ac_2]}{}^{MN}G_{[c_1,bc_2]MN} - {2 \over 5} \delta^b_a G_{[c_1,\ldots c_3]}{}^P G_{[c_1,\ldots c_3]P} ) \; ,
\eqno(5.21) $$
 and ${\cal{E}}_{(QR)}$ by
$$ {\cal{E}}_{(QR)} = {1\over 2} G_{a_1,d}{}^d(G_{a_1,Q}{}^R + G_{a_1, R}{}^Q) - G_{a_1,d}{}^{a_1}(G_{d,Q}{}^R + G_{d, Q}{}^{R})  $$
$$ G_{a_1,Q}{}^N G_{a_1, R}{}^N - G_{a_1,N}{}^Q G_{a_1,N}{}^R + \det e^{1\over2}e_{a_1}{}^{\mu} \partial_{\mu}(G_{a_1,Q}{}^R + G_{a_1,R}{}^Q) $$
$$ + 8 ( G_{[c_1,c_2]QP}G_{[c_1,c_2]RP} + {1\over 5} \delta^Q_R G_{[c_1,c_2]}{}^{NP}G_{[c_1,c_2]NP}) $$
$$ -6 (G_{[c_1,\ldots c_3]Q}G_{[c_1,\ldots c_3]R} + {1\over 5} \delta^Q_R G_{[c_1,\ldots c_3]P} G_{[c_1,\ldots c_3] P}) \; . \eqno(5.22) $$
\par
Since the variation vanishes we find that  $ E_{a}{}^b =0 $ and ${\cal{E}}_{(QR)}=0$. We recognise these as the gravity and scalar equations of motion.
\par
The transformation of the equation of motion for the two form is
$$ \delta {\cal  E}{}^{ a_2a_3M} =   {2\over 3}\Lambda^{[a_2PQ}E{}^{ a_3]RS}\varepsilon_{PQRSM}  $$
$$ + {i \over 3} \Lambda_{cNM}({1\over2} G_{a_1,d}{}^d\varepsilon^{a_1ca_2a_3d_1d_2d_3} - G_{a_1,d}{}^{a_1}\varepsilon^{dca_2a_3d_1d_2d_3} $$
$$ - 2G_{a_1,d}{}^{[a_2|}\varepsilon^{a_1cd|a_3]d_1d_2d_3} ){ D}{}_{d_1d_2d_3}{}^N $$
$$ + {i \over 3} \varepsilon^{a_1ca_2a_3d_1d_2d_3} \det(e)^{1\over2}e_{a_1}{}^{\mu}\partial_{\mu}(\Lambda_{cNM}{ D}{}_{d_1d_2d_3}{}^{N}) \; .
\eqno(5.23) $$
\par
Our final task is the vary the gravity and scalar equations of motion that we have just found. We begin with the gravity equation (5.21).
The first step is to carry out the procedure similar to that at the end of section four but now we do it for the spin connection. Namely we add
terms to the spin connection such that its variation under $I_c(E_{11})$ leads to terms in which the $l_1$ and $E_{11}$ indices on the Cartan forms are antisymmetrised. this is achieved by the object
$$ \det(e)^{1\over2}\Omega_{c,ab} \equiv  \det(e)^{1\over 2} \omega_{c,ab} - {2\over 5}\delta^b_c G^{MN}{}_{,aMN} + {2\over 5}\delta^a_c G^{MN}{}_{,bMN} \; ,  \eqno(5.24) $$
whose  transformation is given by
$$ \delta (  \det(e)^{1\over2}\Omega_{c,ab} ) = - 2\Lambda^{cMN}G_{[a,b]MN} - 2 \Lambda^{bMN}G_{[a,c]MN} + 2 \Lambda^{aMN}G_{[b,c]MN} $$
$$ + {4 \over 5} \delta^b_c \Lambda^{dMN}G_{[a,d]MN} - {4 \over 5} \delta^a_c \Lambda^{dMN}G_{[b,d]MN} \; . \eqno(5.25) $$
We then replace $\omega$ with  $\Omega$ in $R_a{}^b$ to find
$$ {\cal R}_a{}^b = \partial_{\mu}\Omega_{\nu,}{}^{ab} e_b{}^{\nu} - \partial_{\nu}\Omega_{\mu,}{}^{ab}e_b{}^{\nu} + \Omega_{\mu,}{}^{a}{}_c\Omega{}_{\nu,}{}^{cb}e_b{}^{\nu} - \Omega_{\nu,}{}^a_c \Omega_{\mu,}{}^{cb}e_b{}^{\nu} \; .
 \eqno(5.26) $$
\par
As explained above when we vary the gravity equation of motion we also find the terms in this equation of motion that
are first order in derivatives with respect to the level one coordinates, the $l_1$ terms. The result is given by
$$ {\cal E}_{a}{}^b \equiv \det(e) {\cal R}_a{}^b - G_{a,(MN)}G_{b,(MN)} $$
$$ - 2 (2 G_{[c,a]}{}^{MN}G_{[c,b]MN} - {1\over 5} \delta^b_a G_{[c_1,c_2]}{}^{MN}G_{[c_1,c_2]MN}) $$
$$ - 3(3 G_{[c_1,ac_2]}{}^{MN}G_{[c_1,bc_2]MN} - {2 \over 5} \delta^b_a G_{[c_1,\ldots c_3]}{}^P G_{[c_1,\ldots c_3]P} ) $$
$$ - 2 \partial^{MN}G_{[b,a]MN} - 4 G^{NP}{}_{,P}{}^MG_{[b,a]MN} - {2\over 5} G^{MN}{}_{,aMN}\omega_{d,}{}^{bd} + {2\over 5} G^{MN}{}_{,dMN}\omega_{a,}{}^{bd}$$
$$ + 2 G^{MN}{}_{,b}{}^cG_{[c,a]MN} + 2 G^{MN}{}_{,a}{}^cG_{[c,b]MN} $$
$$ - 4G_{b,(MP)}G^{PN}{}_{,aMN} - 4G_{a,(MP)}G^{PN}{}_{,bMN} =0 \; .
\eqno(5.27) $$
Although the  ${\cal R}_a{}^b$ we introduced in equation (5.26) is not symmetric in  $a$ and  $b$ one can verify that $ {\cal E}_{a}{}^b$
is symmetric in  $a$ and  $b$ including the $l_1$ terms that it contains.
\par
The transformation of $ {\cal E}_{a}{}^b$ is given by
$$ \delta {\cal E}_a{}^b = - 2\Lambda_{aMN}E{}^{ bMN} - 2 \Lambda^{bMN}E{}_{ aMN} + {4 \over 5} \delta_a^b\Lambda_{cMN}E{}^{ cMN} $$
$$  -3i \Lambda_{dNP}\varepsilon_{c_1dac_2d_1d_2d_3}G_{[c_1,bc_2]}{}^P{ D}{}_{d_1d_2d_3}{}^{N}  $$
$$  -3i \Lambda_{dNP}\varepsilon_{c_1dbc_2d_1d_2d_3}G_{[c_1,ac_2]}{}^P{ D}{}_{d_1d_2d_3}{}^{N}   $$
$$  +{6 i\over 5}  \delta^b_a \Lambda_{dNP}\varepsilon_{c_1dc_2c_3d_1d_2d_3}G_{[c_1,c_2c_3]}{}^P{D}{}_{d_1d_2d_3}{}^{N}  \; .   \eqno(5.28)$$
\par
Similarly we can find the transformation of the scalar equation of equation (5.22). We find that the equation of motion becomes
$$ {\cal{E}}_{(QR)} \equiv  {1\over 2} G_{a_1,d}{}^d(G_{a_1,Q}{}^R + G_{a_1, R}{}^Q) - G_{a_1,d}{}^{a_1}(G_{d,Q}{}^R + G_{d, Q}{}^{R})  $$
$$ G_{a_1,Q}{}^N G_{a_1, R}{}^N - G_{a_1,N}{}^Q G_{a_1,N}{}^R + \det e^{1\over2}e_{a_1}{}^{\mu} \partial_{\mu}(G_{a_1,Q}{}^R + G_{a_1,R}{}^Q) $$
$$ + 8 ( G_{[c_1,c_2]QP}G_{[c_1,c_2]RP} + {1\over 5} \delta^Q_R G_{[c_1,c_2]}{}^{NP}G_{[c_1,c_2]NP}) $$
$$ -6 (G_{[c_1,\ldots c_3]Q}G_{[c_1,\ldots c_3]R} + {1\over 5} \delta^Q_R G_{[c_1,\ldots c_3]P} G_{[c_1,\ldots c_3] P}) $$
$$ - {8 \over 5} \delta^R_Q \partial_{\mu}(\det(e)^{1\over 2}G^{PN}{}_{,\mu PN})) $$
$$ - 2G_{a_1,d}{}^d G^{PR}{}_{,a_1PQ} + G^{MN}{}_{, dMN} G_{d,Q}{}^{R} + G_{a_1,d}{}^{a_1}G^{PR}{}_{,dPQ} $$
$$  - 4 G_{a_1,R}{}^NG^{PN}{}_{,a_1PQ} + 4 G_{a_1,N}{}^Q G^{PR}{}_{,a_1PN} - 4\det(e)^{1\over 2}e_{a_1}{}^{\mu}\partial_{\mu}(G^{PR}{}_{,a_1QN}) $$
$$ - 2G_{a_1,d}{}^d G^{PQ}{}_{,a_1PR} + G^{MN}{}_{, dMN} G_{d,R}{}^{Q} + G_{a_1,d}{}^{a_1}G^{PQ}{}_{,dPR} $$
$$  - 4 G_{a_1,Q}{}^NG^{PN}{}_{,a_1PR} + 4 G_{a_1,N}{}^R G^{PQ}{}_{,a_1PN} - 4\det(e)^{1\over 2}e_{a_1}{}^{\mu}\partial_{\mu}(G^{PQ}{}_{,a_1RN})=0 \; ,
\eqno(5.29) $$
and its transformation is given by
$$ \delta {\cal E}_{(QR)} \rightarrow 8 \Lambda_{cPR}E{}_{ cPQ} + 8 \Lambda_{cPQ}E{}_{ c PR} -{16 \over 5} \delta_{QR} \Lambda_{cPN}E{}^{ c  PN} $$
$$ + 2i \Lambda_{d PR} \varepsilon^{dd_1d_2d_3c_1c_2c_3}G_{[c_1,c_2c_3]Q}{D}{}_{d_1d_2d_3}{}^P $$
$$ + 2i  \Lambda_{d PQ} \varepsilon^{dd_1d_2d_3c_1c_2c_3}G_{[c_1,c_2c_3]R}{ D}{}_{d_1d_2d_3}{}^P $$
$$ - {4i \over 5}  \delta^R_Q \Lambda_{d PN} \varepsilon^{dd_1d_2d_3c_1c_2c_3}G_{[c_1,c_2c_3]N}{ D}{}_{d_1d_2d_3}{}^P   \; .  
 \eqno(5.30) $$
\par
Using the symmetries of the non-linear realisation we have found a set of equations that transform into each other. These are the equations of motion for the graviton (5.27), scalar (5.29), one form (5.18), and two form (5.19) in seven  dimensions
If we truncate the equations so that they only contain derivatives with respect to the usual coordinates of spacetime then these equations are those of seven dimensional maximal supergravity as found in reference  [14], once we discard terms with derivatives with respect to the level 1 generalised coordinates.
\medskip
{\bf   6 First order duality relations}
\medskip

\par
In this section we will derive  the duality relations  which are first order in derivatives and also find their variations. In addition to those we discussed in section five we will find the duality relations that relate the  graviton to the dual graviton and  the scalar fields to the  dual scalar fields.   We begin by recalling, from section five,  the transformation of the duality relation which relates the two form to the three form and had the form 
$${  D}{}_{a_1a_2a_3}{}^{ M} \equiv  {\cal G}_{a_1a_2a_3}{}^M + {i \over 3} \varepsilon_{a_1a_2a_3}{}^{b_1\ldots b_4} {\cal G}_{b_1\ldots b_4 M }=0 \; , \eqno(6.1) $$
and whose variation was given by
$$
\delta { D}{}_{a_1a_2a_3 M} = \varepsilon^{PQRSM}\Lambda_{[a_2PQ}{D}_{a_1a_3]RS} + {i \over 3} \varepsilon_{a_1a_2a_3}{}^{b_1\ldots b_4} \Lambda_{b_2 NM}{ D}_{b_1b_3b_4}{}^N \; .
 \eqno(6.2) $$
We observe that it transforms into itself and the 1-form duality relation.
\par
In section five, we also transformed the duality relation which relates the one form to the four form  but we did not include the terms in the variation  that contained  the scalar fields,  graviton or their dual  fields which have five spacetime indices and so are at level five. Carrying  out the variation of this duality relation including these additional terms gives the result 
$$ \delta { {\cal D}}_{a_1a_2MN}=  - \Lambda_{[a_2 P[N}{ D}_{a_1], (M] P)} + \Lambda_{aMN}{ { D}}_{b,[a_1a_2]}    - \Lambda^{bMN}{ {D}}_{b,[a_1a_2]}
$$
$$
  -{3\over 2} \varepsilon_{MNPQR}\Lambda_{bQR}{D}_{a_1ba_2}{}^P  + 4i \Lambda_{[a_2}{}_{P[N|} \varepsilon_{a_1]}{}^{b_1\ldots b_6} {\cal G}_{b_1,b_2\ldots b_6 [P |M]]}  
\eqno(6.3) $$
where
$$ { { D}}_{b,[a_1a_2]} \equiv \omega_{b, a_1a_2 }\det(e)^{1\over2} + i \varepsilon_{a_1a_2}{}^{b_1\ldots b_5}{\cal G}_{[b_1,\ldots b_5],b} \; ,
\eqno(6.4) $$
$$
{ D}_{a, (M N)} \equiv 2 G_{a,(MN)} - 4i  \varepsilon_{a}{}^{b_1\ldots b_6}{\cal G}_{b_1,\ldots b_6 (MN)} \; .
\eqno(6.5) $$
and
$$ {\cal D}_{a_1a_2MN} \equiv {\cal G}_{a_1a_2MN} - {i\over 2} \varepsilon_{a_1a_2}{}^{b_1\ldots b_5} {\cal G}_{b_1\ldots b_5}{}^{MN} + G^{MN}{}_{,[a_1a_2]} = 0 \; .
\eqno(6.6) $$
When carrying out the variation to include the extra terms we find that the duality relation of equation (5.1) becomes modified by an $l_1$ term following the procedure  explained earlier in this paper.
\par
As ${ D}_{a_1a_2 M N} =0 $,  its variation under $I_c(E_{11})$ implies that
$$ 
{D}_{b,[a_1a_2]} \; {\dot =} \; 0 \; , \quad  { D}_{a, (M N)} = 0 \; ,
 \eqno(6.7) $$
which are the the duality relation of the scalar field and the graviton respectively, as well as the equation
$$ G_{[a_1, a_2 \ldots  a_6] [M N]} \; \dot= \; 0 \; .
\eqno(6.8) $$
The dots above the equal signs signify that the equations only hold modulo certain local symmetries as explained in references [15, 8, 16, 17]. 
Equation (6.8)  sets the field strength for the dual scalar field $A_{a_1 \ldots  a_5 [M N]} $ to be zero and so this field is pure gauge and can be removed by the gauge transformation that exists at this level. 
\par
We now transform the new duality relations we have found in equation (6.7) using the variation given in section four.   Varying the scalar field duality relations of  equation (6.5) we find the result
$$
\delta {\cal D}_{a, (MN)} = 8 \Lambda_{cPN}{ D}^{ac PM} + 8 \Lambda_{cPM}{ D}^{ac PN} - {16 \over 5} \delta_N^M \Lambda_{cPQ}D^{ac PQ} +\ldots 
\eqno(6.9) $$
While the variation of the gravity duality relation is given by
$$
 \delta {\cal D}_{b,[a_1a_2]} = - 2\Lambda^{bMN}{ D}_{a_1a_2 MN} + {4\over5} \Lambda^{dMN} \delta^b_{[a_2}{ D}_{a_1]dMN} $$ $$ +{4\over 6!5} \Lambda^{dMN}\delta^b_{[a_2}\varepsilon_{a_1]}{}^{d_1\ldots d_6}{ D}_{d_1\ldots d_6,d MN} $$
$$ - {4\over 6!}\Lambda_{[a_2 MN}\varepsilon_{a_1]}{}^{d_1\ldots d_6}{ D}_{d_1\ldots d_6,b MN} + \partial_{b} {\tilde \Lambda}_{a_1a_2} \;  ,
\eqno(6.10) $$
where
$$ { D}_{a_1\ldots a_6,b MN} \equiv 2(3!)^2i G_{[a_1,\ldots a_6],b MN} + \varepsilon_{a_1\ldots a_6}{}^d{\cal G}_{[d,b] MN}=0 \; ,
\eqno(6.11) $$
and
$$
  \partial_{b} {\tilde \Lambda}_{a_1a_2} = -{i\over 25} \varepsilon_{a_1a_2}{}^{b_1\ldots b_5}(\Lambda_{b_1MN}{\cal G}_{b,b_2\ldots b_5}{}^{MN} + \Lambda^{cMN} G_{b,b_1\ldots b_5,cMN})  \; .
\eqno(6.12) $$
\par
In equation (6.11) we have a new duality relation which involves a level six field. This term is analogous to the duality relation  connecting  the $A_{a_1a_2a_3}$ field to the  $A_{b_1\ldots b_9, a_1a_2a_3}$ in eleven dimensions as discussed in reference [16]. In equation (6.12) we find the Lorentz transformations which should be expected as we are varying a duality relation that only holds modulo Lorentz transformations, also like the situtation in reference [15].
\par
When transforming the scalar and graviton duality relations we find they are modified by $l_1$ terms. Including these terms  the duality relations used in the variation are
$$
 {  { \cal D}}_{b,[a_1a_2]} \equiv \Omega_{b,[a_1a_2]}\det(e)^{1\over2} + i \varepsilon_{a_1a_2}{}^{b_1\ldots b_5}{\cal G}_{[b_1,\ldots b_5],b} \; ,
\eqno(6.13)$$
and
$$ 
  {\cal D}_{a, (M N)} \equiv  2 G_{a,(MN)} - 4i  \varepsilon_{a}{}^{b_1\ldots b_6}{\cal G}_{b_1,\ldots b_6 (MN)} - 4 G^{PN}{}_{,aPM} - 4G^{PM}{}_{,aPN} + {4 \over 5} \delta^M_NG^{PQ}{}_{,aPQ} \; ,
  \eqno(6.14) $$

In the first relation we have replaced $ \omega_{b,[a_1a_2]}$ by $ \Omega_{b,[a_1a_2]}$ which includes the required $l_1$ terms as we did in   equation (5.24). 
\par
The factors of $i$ that appear in the  above relations can be removed from all the equations in this paper by some field redefinitions. We take the parameter $ \Lambda_{aNM}\to i  \Lambda_{aNM}$, change  the fields with an odd number of Lorentz indices by a factor of $i$, leave the fields with an even number of indices the same and finally put a factor of $i$ with $\partial^{MN}$.
\par
In this section we have found that the fields up to level five satisfy duality relations that are first order in derivatives vary under  $I_c(E_{11})$ into themselves as well as relations involving level six fields. It would be interesting to carry out the calculation to determine the later duality relations which involve the fields responsible for the gauged seven dimensional supergravities. It would also be interesting to derive all the second order equations of motion of section five from the first order duality relation as was done in eleven dimensions in [17]. However, as was spelt out there this is a subtle procedure as one must correctly take account of the fact that some of these relations hold modulo certain transformations.


\medskip
{\bf   7 Conclusion}
\medskip
In this paper we have constructed, at low levels,  the non-linear realisation of the semi-direct product of $E_{11}$ with its vector representation in seven dimensions. The resulting dynamical equations follow essentially uniquely from the $E_{11}$ Dynkin diagram once we delete node seven and take the corresponding decomposition. These equations agree with those of seven dimensional supergravity if only keep  the fields used in the usual description of seven dimensional supergravity as well as  keep only derivatives with respect to the usual spacetime coordinates. It has been proposed  that one can,  by taking different decompositions of $E_{11}$, find all the massless maximal supergravities from the  the non-linear realisation of the semi-direct product of $E_{11}$ with its vector representation by taking the different decompositions. It is good to see how this works in detail  in seven dimensions in this paper. 
\par
Systematically extending this calculation to level six we will find the dynamics for the six form gauge fields and it would be good to examine in detail how these lead to all the gauged supergravities in seven dimensions as indeed  it did in ten dimensions [16]. We hope to carry out this calculation in a future paper. 
\medskip
{\bf Appendix A The generators of $E_{11}$ decomposed into representations of of $I_c(E_{11})$ }
\medskip
The $E_{11}$ algebra can be split into those that are even under the action of the Cartan involution and those that are odd. The former are by definition the algebra $I_c(E_{11})$ and the latter belong to a representation of $I_c(E_{11})$,  their commutators belong to $I_c(E_{11})$ and they are given by
$$
T^a{}_b = R^a{}_b + R^b{}_a, \; ; \ \
T^M{}_N = R^M{}_N + R^N{}_M  \; ;\ \
T^{aMN} = R^{aMN} + R_{aMN} \; ;
$$
$$
T^{a_1a_2}{}_M = R^{a_1a_2}{}_M - R_{a_1a_2}{}^M
\; ; \ \
T^{a_1a_2a_3M} = R^{a_1a_2a_3M} + R_{a_1a_2a_3M} \; ; \; \ldots  \; .
\eqno(A.1)$$
\par
In this appendix we compute the commutators between  these generators. Firstly, the commutators of the level 0 generators are given by
$$
[T^a{}_b, T^c{}_d] = \delta^c_bJ^a{}_d + \delta^c_aJ^b{}_d - \delta^a{}_dJ^c{}_b - \delta^b_dJ^c{}_a  \; ,
\eqno(A.2)
$$
$$
[T^M{}_N, T^P{}_Q] = \delta^P_NS^M{}_Q + \delta^P_MS^N{}_Q - \delta^M_QS^P{}_N - \delta^N_QS^P{}_M   \; ,
\eqno(A.3)
$$
$$
[T^a{}_b, T^M{}_N] = 0   \; .
\eqno(A.4)
$$
Then the level 0  generators with the positive level generators are given by
$$
[T^a{}_b, T^{cMN}] = \delta^c_bS^{aMN} + \delta^c_aS^{bMN}   \; ,
\eqno(A.5)
$$
$$
[T^a{}_b, T^{cd}{}_M] = 2\delta_b^{[c}S^{|a|d]}{}_M + 2 \delta_a^{[c}S^{|b|d]}{}_M   \; ,
\eqno(A.6)
$$
$$
[T^a{}_b, T^{c_1c_2c_3M}] = 3\delta_b^{[c_1}S^{|a|c_2c_3]M} + 3\delta_a^{[c_1}S^{|b|c_2c_3]M}   \; ,
\eqno(A.7)
$$
$$
[T^M{}_N, T^{aPQ}] = 2\delta_N^{[P}S^{a|M|Q]} + 2\delta_M^{[P}S^{a|N|Q]} - {4\over5}\delta^M_NS^{aPQ}   \; ,
\eqno(A.8)
$$
$$ [T^M{}_N, T^{ab}{}_P] = -\delta^M_PS^{ab}{}_N - \delta^N_PS^{ab}{}_M + {2\over 4}\delta^M_NS^{ab}{}_P   \; ,
\eqno(A.9)
$$
$$
[T^M{}_N, T^{a_1a_2a_3P}] = \delta^P_NS^{a_1a_2a_3M} + \delta^P_MS^{a_1a_2a_3N} - {2\over5}\delta^M_NS^{a_1a_2a_3P}   \; ,
\eqno(A.10)
$$
\par
The commutators of the positive level generators  are given by
$$
[T^{aMN}, T^{bPQ}] = \varepsilon^{MNPQR}S^{ab}{}_R + \delta^a_b\delta^{[M}_{[P}S^{N]}{}_{Q]} + 2\delta^{MN}_{PQ}J^a{}_b   \; ,
\eqno(A.11)
$$
$$
[T^{aMN}, T^{b_1b_2}{}_P] = \delta_P^{[M}S^{ab_1b_2N]} -\varepsilon^{MNPQR}\delta_a^{[b_1}S^{b_2]QR}  \; ,
\eqno(A.12)
$$
$$
[T^{aMN}, T^{b_1b_2b_3P}] = -12\delta_a^{[b_1}\delta^P_{[M}S^{b_2b_3]}{}_{N]}  \; ,
\eqno(A.13)
$$
$$
[T^{a_1a_2}{}_M, T^{b_1b_2}{}_N] = -{1\over 2}\delta^{a_1a_2}_{b_1b_2}S^M{}_N - 4\delta^M_N\delta^{[a_1}_{[b_1}J^{a_2]}{}_{b_2]}  \; ,
\eqno(A.14)
$$
$$
[T^{a_1a_2}{}_M, T^{b_1b_2b_3N}] = 12\delta_{a_1a_2}^{[b_1b_2}S^{b_3]MN}  \; ,
\eqno(A.15)
$$
$$[T^{a_1a_2a_3M}, T^{b_1b_2b_3N}] = -4!\delta^{a_1a_2a_3}_{b_1b_2b_3}S^M{}_N - 2\cdot (3!)^2\delta^M_N\delta^{[a_1a_2}_{[b_1b_2}J^{a_3]}{}_{b_3]}  \; .
\eqno(A.16)
$$
\par
We next give the commutators with those of the generators of the vector representation. At level 0, we find
$$
[T^a{}_b, P_c] = - \delta^a_cP_b - \delta^b_cP_a + \delta^a_bP_c  \; ,
\eqno(A.17)
$$
$$
[T^a{}_b, Z^{MN}] = \delta^a_bZ^{MN}  \; ,
\eqno(A.18)
$$
$$
[T^a{}_b, Z^c{}_M] = \delta^c_b Z^a{}_M + \delta^c_aZ^b{}_M + \delta^a_bZ^c{}_M  \; ,
\eqno(A.19)
$$
$$
[T^{a}{}_b, Z^{a_1a_2M}] = 2\delta_b^{[a_1}Z^{|a|a_2]M} + 2\delta_a^{[a_1}Z^{|b|a_2]M} + \delta^a_bZ^{a_1a_2M}  \; ,
\eqno(A.20)
$$

$$
[T^M{}_N, P_a] = 0  \; ,
\eqno(A.21)
$$
$$
[ T^M{}_N, Z^{PQ}] = 2\delta_N^{[P}Z^{|M|Q]} + 2\delta_M^{[P}Z^{|N|Q]} - {4\over 5}\delta^M_NZ^{PQ}   \; ,
\eqno(A.22)
$$
$$
[T^M{}_N, Z^{a}{}_P] = -\delta^M_P Z^a{}_N - \delta^N_PZ^a{}_M +{2\over5}\delta^M_NZ^a{}_P   \; ,
\eqno(A.23)
$$
$$
[T^M{}_N, Z^{a_1a_2P}] = \delta^P_NZ^{a_1a_2M} + \delta^P_MZ^{a_1a_2N} - {2\over 5}\delta^M_NZ^{a_1a_2P}  \; .
\eqno(A.24)
$$
The commutators of the level 1 generators with the $l_1$ representation are
$$
[T^{aMN}, P_b] = \delta^a_bZ^{MN}  \; ,
\eqno(A.25)
$$
$$
[T^{aMN}, Z^{PQ}] = - \varepsilon^{MNPQR}Z^a{}_R + 2\delta^{PQ}_{MN}P_a   \; ,
\eqno(A.26)
$$
$$
[T^{aMN}, Z^b{}_P] = 2\delta_P^{[M}Z^{abN]} - {1\over2}\delta^a_b\varepsilon_{MNPQR}Z^{QR}  \; ,
\eqno(A.27)
$$
$$[T^{aMN}, Z^{b_1b_2P}] = -4\delta^P_{[M}\delta_a^{[b_1}Z^{b_2]}{}_{N]}  \; .
\eqno(A.28)
$$
Then the commutators of the level 2 generators with the $l_1$ representation are
$$
[T^{a_1a_2}{}_M, P_b] = 2\delta_b^{[a_1}Z^{a_2]}{}_M   \; ,
\eqno(A.29)
$$
$$
[T^{a_1a_2}{}_M, Z^{NP}] = 2\delta_M^{[N}Z^{a_1a_2P]}   \; ,
\eqno(A.30)
$$
$$
[T^{a_1a_2}{}_M, Z^b{}_N] = -2 \delta^M_N\delta^b_{[a_1}P_{a_2]}   \; ,
\eqno(A.31)
$$
$$
[T^{a_1a_2}{}_M, Z^{b_1b_2N}] = 2\delta^{b_1b_2}_{a_1a_2}Z^{MN}   \; .
\eqno(A.32)
$$
Finally, the commutators of the level 3 generators with the $l_1$ representation are
$$
[T^{a_1a_2a_3M}, P_b] = - 6\delta_b^{[a_1}Z^{a_2a_3]M}   \; ,
\eqno(A.33)
$$
$$
[T^{a_1a_2a_3M}, Z^{NP}] = 0   \; ,
\eqno(A.34)
$$
$$
[T^{a_1a_2a_3M}, Z^b{}_N] = 0   \; ,
\eqno(A.35)
$$
$$
[T^{a_1a_2a_3M},Z^{b_1b_2N}] = 12\delta^N_M\delta^{b_1b_2}_{[a_1a_2}P_{a_3]}   \; .
\eqno(A.36)
$$
\par
Finally we give the commutators of the even generators, that is, $I_c(E_{11})$ with the odd generators. At level zero we have
$$
[J^a{}_b, T^c{}_d] = \delta^c_bT^a{}_b + \delta^b_dT^c{}_a - \delta^c_aT^b{}_d - \delta^a_dT^c{}_b   \; ,
\eqno(A.37)
$$
$$
[J^a{}_b, T^M{}_N] = 0   \; ,
\eqno(A.38)
$$
$$
[S^M{}_N, T^c{}_d] = 0   \; ,
\eqno(A.39)
$$
$$
[S^M{}_N, T^P{}_Q] = \delta^P_NT^M{}_Q + \delta^N_QT^P{}_M- \delta^P_MT^N{}_Q - \delta^M_QT^P{}_N   \; .
\eqno(A.40)
$$
The level 0 even with odd generators are given by
$$
[J^a{}_b, T^{cMN}] = \delta^c_bT^{aMN} - \delta^c_aT^{bMN}   \; ,
\eqno(A.41)
$$
$$
[J^a{}_b, T^{cd}{}_M] = 2\delta_b^{[c}T^{|a|d]}{}_M - 2 \delta_a^{[c}T^{|b|d]}{}_M   \; ,
\eqno(A.42)
$$
$$
[J^a{}_b, T^{c_1c_2c_3M}] = 3\delta_b^{[c_1}T^{|a|c_2c_3]M} - 3\delta_a^{[c_1}T^{|b|c_2c_3]M}  \; ,
\eqno(A.43)
$$
$$
[S^M{}_N, T^{aPQ}] = 2\delta_N^{[P}T^{a|M|Q]} - 2\delta_M^{[P}T^{a|N|Q]}  \; ,
\eqno(A.44)
$$
$$ [S^M{}_N, T^{ab}{}_P] = -\delta^M_PT^{ab}{}_N + \delta^N_PT^{ab}{}_M   \; ,
\eqno(A.35)
$$
$$
[S^M{}_N, T^{a_1a_2a_3P}] = \delta^P_NT^{a_1a_2a_3M} - \delta^P_MT^{a_1a_2a_3N}   \; ,
\eqno(A.46)
$$
and the level 1 even generators with the odd generators
$$
[S^{aMN}, T^b{}_c] = - \delta^a_cT^{bMN} - \delta^a_bT^{cMN}  \; ,
\eqno(A.47)
$$
$$
[S^{aMN}, T^P{}_Q] = -2\delta_Q^{[M}T^{|a|P|N]} - 2\delta_P^{[M}T^{a|Q|N]}   \; ,
\eqno(A.48)
$$
$$
[S^{aMN}, T^{bPQ}] = \varepsilon^{MNPQR}T^{ab}{}_R + \delta^a_b\delta^{[M}_{[P}T^{N]}{}_{Q]} + 2\delta^{MN}_{PQ}T^a{}_b - {2\over 5}\delta^{MN}_{PQ}\delta^a_bT^c{}_c  \; ,
\eqno(A.49)
$$
$$
[S^{aMN}, T^{b_1b_2}{}_P] = \delta_P^{[M}T^{ab_1b_2N]} +\varepsilon^{MNPQR}\delta_a^{[b_1}T^{b_2]QR}  \; ,
\eqno(A.50)
$$
$$
[S^{aMN}, T^{b_1b_2b_3P}] = 12\delta_a^{[b_1}\delta^P_{[M}T^{b_2b_3]}{}_{N]} \; .
\eqno(A.51)
$$
\noindent
Level 2 even generators with the odd generators are
$$
[S^{a_1a_2}{}_M, T^b{}_c] = - 2\delta_c^{[a_1}T^{|b|a_2]}{}_M - 2\delta_b^{[a_1}T^{|c|a_2]}{}_M   \; ,
\eqno(A.52)
$$
$$
[S^{a_1a_2}{}_M, T^N{}_P] = - \delta^N_MT^{a_1a_2}{}_P - \delta^P_MT^{a_1a_2}{}_N + {2\over5}\delta^N_PT^{a_1a_2}{}_M  \; ,
\eqno(A.53)
$$
$$
[S^{a_1a_2}{}_M, T^{bNP}] = -\delta_M^{[N}T^{ba_1a_2P]} + \varepsilon_{MNPQR}\delta_b^{[a_1}T^{a_2]QR}  \; ,
\eqno(A.54)
$$
$$
[S^{a_1a_2}{}_M, T^{b_1b_2}{}_N] = -{1\over 2}\delta^{a_1a_2}_{b_1b_2}T^M{}_N - 4\delta^M_N\delta^{[a_1}_{[b_1}T^{a_2]}{}_{b_2]} + {4\over5}\delta^N_M\delta^{a_1a_2}_{b_1b_2}T^c{}_c  \; ,
\eqno(A.45)
$$
$$
[S^{a_1a_2}{}_M, T^{b_1b_2b_3N}] = - 12\delta_{a_1a_2}^{[b_1b_2}T^{b_3]MN}  \; .
\eqno(A.56)
$$
\noindent
Finally, the level 3 even generators with the odd generators are
$$[S^{a_1a_2a_3M}, T^b{}_c] = - \delta_c^{[a_1}T^{|b|a_2a_3]M} - \delta_b^{[a_1}T^{|c|a_2a_3]M}   \; ,
\eqno(A.57)
$$
$$[S^{a_1a_2a_3M}, T^N{}_P] = -\delta_P^MT^{a_1a_2a_3N} - \delta^M_NT^{a_1a_2a_3P} - {2\over5}\delta^N_PT^{a_1a_2a_3M}  \; ,
\eqno(A.48)
$$
$$[S^{a_1a_2a_3M}, T^{bNP}] = 12\delta_b^{[a_1}\delta^M_{[N}T^{a_2a_3]}{}_{P]}   \; ,
\eqno(A.59)
$$
$$[S^{a_1a_2a_3M}, T^{b_1b_2}{}_N] = - 12\delta_{b_1b_2}^{[a_1a_2}T^{a_3]MN}   \; ,
\eqno(A.60)
$$
$$[S^{a_1a_2a_3M}, T^{b_1b_2b_3N}] = -4!\delta^{a_1a_2a_3}_{b_1b_2b_3}T^M{}_N $$
$$- 2\cdot (3!)^2\delta^M_N\delta^{[a_1a_2}_{[b_1b_2}T^{a_3]}{}_{b_3]} + {2\over5}\cdot (3!)\delta^M_N\delta^{a_1a_2a_3}_{b_1b_2b_3}T^c{}_c  \; .
\eqno(A.61)
$$

\medskip

{\bf Acknowledgements}
Peter West  wishes to thank the SFTC for support from Consolidated grants number ST/J002798/1 and ST/P000258/1 and MIchaella Pettit for an EPSRC PhD grant no EP/M50788X/1.


\medskip
{\bf {References}}
\medskip
\item{[1]} P. West, {\it $E_{11}$ and M Theory}, Class. Quant.
Grav.  {\bf 18}
(2001) 4443, {\tt arXiv:hep-th/ 0104081};
\item{[2]} P. West, {\it $E_{11}$, SL(32) and Central Charges},
Phys. Lett. {\bf B 575} (2003) 333-342, {\tt hep-th/0307098}
\item{[3]} I. Schnakenburg and  P. West, {\it Kac-Moody 
symmetries of
IIB supergravity}, Phys. Lett. {\bf B517} (2001) 421, hep-th/0107181.
\item{[4]} P. West, {\it The IIA, IIB and eleven dimensional theories
and their common
$E_{11}$ origin}, Nucl. Phys. B693 (2004) 76-102, hep-th/0402140.
\item{[5]}  F.  Riccioni and P. West, {\it
The $E_{11}$ origin of all maximal supergravities},  JHEP {\bf 0707}
(2007) 063;  arXiv:0705.0752.
\item{[6]}  F. Riccioni and P. West, {\it E(11)-extended spacetime
and gauged supergravities},
JHEP {\bf 0802} (2008) 039,  arXiv:0712.1795.
\item{[7]} A. Tumanov and P. West, {\it E11 must be a symmetry of strings and branes },  Phys. Lett. {\bf  B759 } (2016),  663, arXiv:1512.01644.
\item{[8]} A. Tumanov and P. West, {\it E11 in 11D}, Phys.Lett. B758 (2016) 278, arXiv:1601.03974.
\item{[9]} P. West, {\it A brief review of E theory}, Proceedings of Abdus Salam's 90th  Birthday meeting, 25-28 January 2016, NTU, Singapore, Editors L. Brink, M. Duff and K. Phua, World Scientific Publishing and IJMPA, {\bf Vol 31}, No 26 (2016) 1630043,  arXiv:1609.06863.
\item{[10]} F.  Riccioni,  D.  Steele and P. West, {\it The E(11) origin of all maximal supergravities - the hierarchy of field-strengths}
  JHEP {\bf 0909} (2009) 095, arXiv:0906.1177.
\item{[11]} P. Cook and P. West, {\it Charge multiplets and masses
for E(11)},  JHEP {\bf 11} (2008) 091, arXiv:0805.4451.
\item{[12]} P. West,  {\it $E_{11}$ origin of Brane charges and U-duality
multiplets}, JHEP 0408 (2004) 052, hep-th/0406150.
\item{[13]} P. West, {\it Brane dynamics, central charges and
$E_{11}$}, JHEP 0503 (2005) 077, hep-th/0412336.
\item{[14]} A. Salam and E. Sezguin, {\it SO(4) Gauging of $N=4$ Supergravity in seven dimensions}, Phys. Lett. {\bf 126B} (1983) 295; {\it Maximal extended Supergravity theory in seven dimensions}, Phys. Lett {\bf 118B} (1982) 359.
\item{[15]} P. West, {\it Dual gravity and E11},  arXiv:1411.0920.
\item{[16]} A. Tumanov and and P. West, {\it $E_{11}$,  Romans theory and higher level duality relations}, IJMPA, {\bf Vol 32}, No 26 (2017) 1750023,  arXiv:1611.03369.  
\item{[17]} P. West, {\it On the different formulations of the E11 equations of motion}, Mod.Phys.Lett. A32 (2017) no.18, 1750096,  arXiv:1704.00580.

\bye